\documentclass[footinbib,a4paper,aps,prB,reprint,twocolumn,preprintnumbers,amsmath,amssymb,10pt, superscriptaddress]{revtex4-2}
\usepackage{verbatim}
\usepackage{graphicx}
\usepackage{color}
\usepackage{tikz}
\usepackage[colorlinks=true,citecolor=blue,linkcolor=blue,urlcolor=blue]{hyperref}
\usepackage{braket}
\usepackage[normalem]{ulem}
\usepackage{upgreek}
\usepackage[percent]{overpic}
\usepackage{graphicx,xcolor}
\usepackage{dcolumn}
\usepackage{bm}
\usepackage{mathtools}
\usepackage{physics}
\usepackage[shortlabels]{enumitem}

\newcommand{\ph}{\phantom{\dag}}

\DeclareFontFamily{U}{wncy}{}
\DeclareFontShape{U}{wncy}{m}{n}{<->wncyr10}{}
\DeclareSymbolFont{mcy}{U}{wncy}{m}{n}
\DeclareMathSymbol{\Sh}{\mathord}{mcy}{"58} 

\begin{document}

\title{Phase diagram of Rydberg-dressed atoms on two-leg square ladders: \\ Coupling supersymmetric conformal field theories on the lattice}
\author{Mikheil Tsitsishvili}%
\email{mtsitsis@ictp.it}
\affiliation{The Abdus Salam International Centre for Theoretical Physics (ICTP), Strada Costiera 11, 34151 Trieste,
Italy}
\affiliation{International School for Advanced Studies (SISSA), via Bonomea 265, 34136 Trieste, Italy}
\author{Titas Chanda}
\email{tchanda@ictp.it}
\affiliation{The Abdus Salam International Centre for Theoretical Physics (ICTP), Strada Costiera 11, 34151 Trieste,
Italy}

\author{Matteo Votto}%
\affiliation{The Abdus Salam International Centre for Theoretical Physics (ICTP), Strada Costiera 11, 34151 Trieste,
Italy}
\author{Pierre Fromholz}%
\email{pierre.fromholz@unibas.ch}
\affiliation{The Abdus Salam International Centre for Theoretical Physics (ICTP), Strada Costiera 11, 34151 Trieste,
Italy}
\affiliation{International School for Advanced Studies (SISSA), via Bonomea 265, 34136 Trieste, Italy}

\author{Marcello Dalmonte}
 \affiliation{The Abdus Salam International Centre for Theoretical Physics (ICTP), Strada Costiera 11, 34151 Trieste,
Italy}
\affiliation{International School for Advanced Studies (SISSA), via Bonomea 265, 34136 Trieste, Italy}

\author{Alexander Nersesyan}
 \affiliation{The Abdus Salam International Centre for Theoretical Physics (ICTP), Strada Costiera 11, 34151 Trieste,
Italy}
\affiliation{The Andronikashvili Institute of Physics, 0177 Tbilisi, Georgia}
\affiliation{Ilia State University, 0162 Tbilisi, Georgia}

\begin{abstract}
We investigate the phase diagram of hard-core bosons in two-leg ladders in the presence of soft-shoulder potentials. We show how the competition between local and non-local terms gives rise to a phase diagram with liquid phases with dominant cluster, spin-, and density-wave quasi-long-range ordering. These phases are separated by Berezinskii-Kosterlitz-Thouless, Gaussian, and supersymmetric (SUSY) quantum critical transitions. For the latter, we provide a phenomenological description of coupled SUSY conformal field theories, whose predictions are confirmed by matrix-product state simulations. Our results are motivated by, and directly relevant to, recent experiments with Rydberg-dressed atoms in optical lattices, where ladder dynamics has already been demonstrated, and emphasize the capabilities of these setups to investigate exotic quantum phenomena such as cluster liquids and coupled SUSY conformal field theories.\end{abstract}

\maketitle

\section{Introduction}
\label{sec:intro}

Over the last ten years, ensembles of ground state atoms laser-coupled to Rydberg states and trapped by means of optical potentials have demonstrated impressive capabilities of realizing strongly interacting quantum dynamics under controlled and tunable experimental conditions~\cite{cheneau2012,Jau2015,Faoro:2016aa,Zeiher2016,Bernien2017,Barredo2018}. From the many-body perspectives, the opportunities offered by these settings profit from the combination of very rich interactions properties between Rydberg states, including long-range characters and spatial anisotropy, with a high degree of local control and manipulation~\cite{Balewski2014}. Within such settings, inter-particle interactions are indeed sizeable even at distances of a few microns -- so that, for instance, lattice potentials can be probed at the single site level with minimal experimental effort when compared to conventional cold atoms in optical lattices operating in Hubbard-like regimes.

The dynamics of Rydberg atom arrays in the case of large lattice spacing between sites is often described in terms of Ising or XY spin-$1/2$ models~\cite{deLeseleuc2018,Browaeys2020,scholl2021microwaveengineering}. Starting from the observation of dynamical crystallization reported in~\cite{Pohl2010,cheneau2012}, a plethora of phenomena has been observed, including magnetic phases of one-dimensional (1D) systems~\cite{Keesling:aa}, the first example of a topological phase in a cold atom experiment~\cite{fde_leuc_2019}, and ordered phased  of two-dimensional (2D) arrays~\cite{Ebadi2021,Scholl2021}. One key leitmotif of these experiments is that they are carried out in such a way that Rydberg states are populated, making the dynamics so fast that atomic motion can most often be neglected as it is slower than typical decoherence mechanism such as spontaneous decay of the Rydberg states.
\begin{figure}[!h]
\centering
   \begin{overpic}[width=0.8\linewidth]{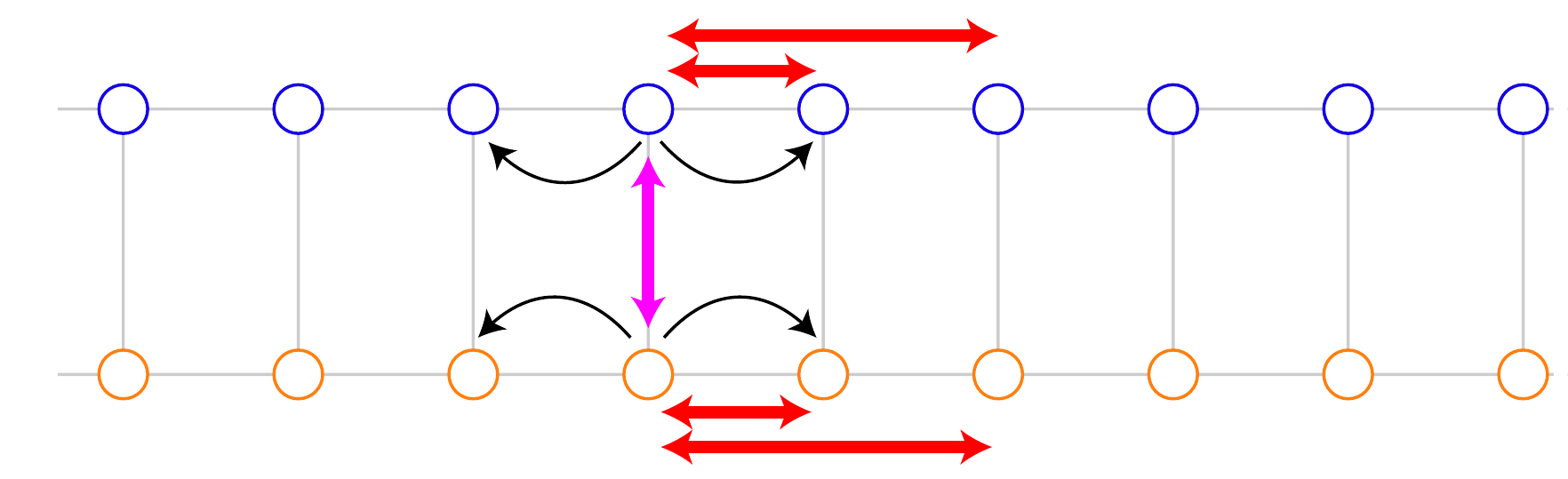}
   \put (-1,26) {{\textbf{(a)}}} \put (0,20) {{$+$}}  \put (0,7) {{$-$}} \put (35,12) {{$t$}} \put (42,13) {{$U$}} \put (34,26) {{$V\lbrace$}}
      \end{overpic}\\ \begin{overpic}[width=0.8\linewidth]{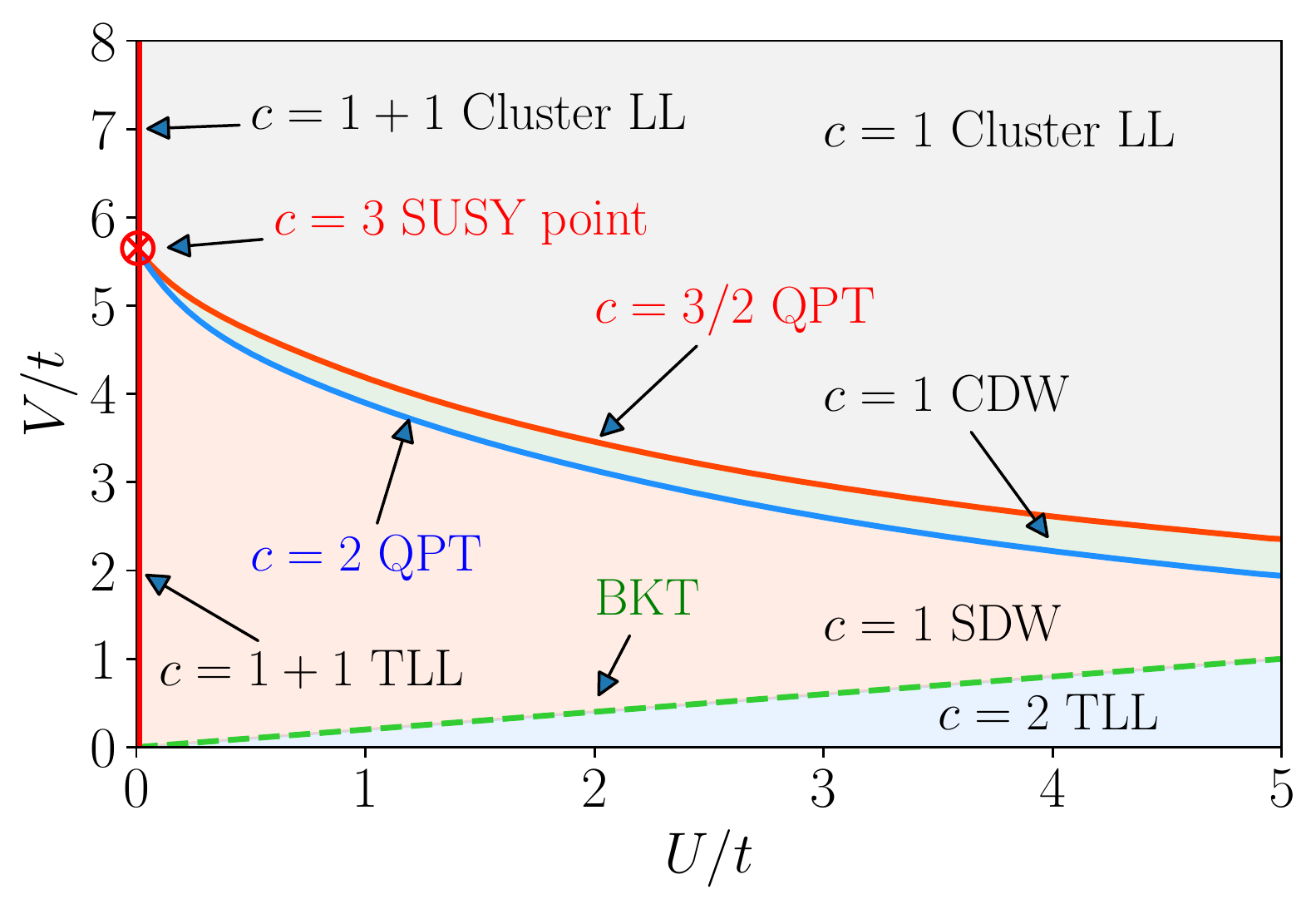}
   \put (-1,66) {{\textbf{(b)}}} 
      \end{overpic} 
  \caption{\label{fig:Ryd} (Color online.) \textbf{(a)} Representation of the soft-shoulder ($V$) Hubbard ($U$) square ladder of Rydberg hard-core bosons. They hop ($t$) within a same leg of the ladder only. We depict a range of 2 for the soft-shoulder interaction. \textbf{(b)} Phase diagram of the model. The weak coupling regime displays a gapless phase composed of two Tomonaga-Luttinger liquids (TLLs) in both charge and spin (mapped from the leg index) sectors (2TLL). Upon increasing $V$, first, the spin sector undergoes a Berezinskii-Kosterlitz-Thouless (BKT) transition (green line) onto a spin-density wave (SDW); then,  a Gaussian transition (blue line) drives the system to a charge-density wave phase (CDW). At large interactions, a $c=3/2$ transition (red line) separates the CDW from a `spin-locked' cluster Luttinger liquid (CLL). The transitions lines at intermediate couplings relate to the effective field theory emerging from the two superposed supersymmetric (SUSY) conformal critical points at $U=0$ (red crossed-dot). At $U=0$, the two legs are independent and display TLL (1+1 TLL) and CLL (1+1 CLL) phases separated by the $c=3$ SUSY point.}
\end{figure}
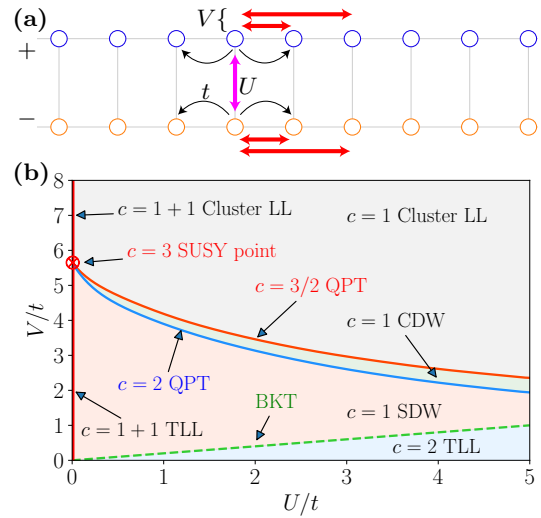

An alternative, relatively unexplored setting is ground state atoms that are only weakly coupled to Rydberg states~\cite{Henkel2010,Pupillo2010,honer2010,Macri2014}: in this regime, up to timescales that are long compared to typical sources of dissipation, the Rydberg state is only virtually populated -- yet, it does have drastic effects in determining the system's coherent dynamics. In particular, it allows to engineer dynamics that is somehow intermediate between conventional Hubbard models, and frozen Rydberg gases: this results into generalized Hubbard models, where interactions combine a long-range, power-law tail (van der Waals), with a short-range plateau, that is of widespread use in classical statistical mechanics, and is often referred to as the soft-shoulder potential~\cite{Browaeys2020}. Even in their simplest instance of a single scalar bosonic field in 1D, these types of interactions lead to exotic critical behavior, including phases where the Luttinger theorem is inapplicable~\cite{Mattioli2013} and supersymmetric quantum critical points~\cite{Dalmonte2015}. The phenomenology is equally rich in 2D systems, where soft-shoulder potentials have been linked to anomalous dynamics and glassy behavior \cite{Angelone_2Ddyn,2016Angelone}. 

Opposite to the aformentioned cases, the transition regime between 1D and 2D systems is poorly understood. The goal of this work is to shed light on the latter, by investigating the ground state phase diagram of soft-shoulder Hubbard models in square ladder geometries (Fig.~\ref{fig:Ryd}\textbf{(a)}). In addition to this theoretical motivation, a recent experiment~\cite{Guardado-Sanchez2021} has realized many-body dynamics of single-species Hubbard models with clustering interactions. In two-leg square ladders, our findings below are, in large part, motivated by such experimental capabilities, and offer a clear theoretical pathway to investigate the effects of frustration within that platform. 

Utilizing a combination of field theoretical approaches based on bosonization~\cite{Gogolin2004,Giamarchi2003}, and numerical simulations based on exact diagonalization~\cite{Sandvik2010} and tensor networks~\cite{schollwock_aop_2011, Orus_aop_2014}, we find that two-leg soft-shoulder Hubbard models support a rich phase diagram (schematically depicted in Fig.~\ref{fig:Ryd}\textbf{(b)}). In the weakly interacting regime, the system supports a fully gapless and a spin-density wave phase, that are well captured via an Abelian bosonization field theory. In the limit of strong intra-chain interactions, the system is effectively described by chains that are subjected to clustering: remarkably, in such limit, despite the already strong nature of fractionalization due to the inapplicability of the Luttinger theorem, finite inter-chain interactions immediately lock clusters, giving rise to a single, `spin-locked' cluster Luttinger liquid. We capture this regime utilizing a phenomenological cluster bosonization field theory~\cite{Haldane1981,Mattioli2013,Dalmonte2015,Gotta2021}.

Close to the (decoupled) single chain critical point, the models we consider offer an almost unique  possibility of investigating the controlled coupling of two supersymmetric (SUSY) conformal field theories~\cite{DIXON1988470,DiFrancesco1997}. This happens at intermediate interactions, where we are unable to construct a microscopically justified quantum field theory. To describe such scenario, we instead propose a phenomenological field theory. The emerging picture is that, for the case of square ladders, the mutual interactions between the SUSY theories lead to the opening of a gapless phase with density-wave quasi-long-range order that is absent in any interaction limit. This phase is separated from the weak coupling phase by a Gaussian transition, and from the strong coupling phase by a conformal critical line with central charge $c=3/2$. We confront these predictions against extensive tensor network simulations, including an analysis of entanglement properties as well as experimentally observable correlation functions.  

The structure of the paper is as follows. In Sec~\ref{sec:ham}, we introduce the model Hamiltonian, its symmetries, and provide an extended qualitative discussion of the phase diagram. In Sec.~\ref{sec:theory}, we discuss the bosonization field theories applicable at weak and strong coupling regimes. In Sec.~\ref{sec:effectiveTheory}, we present a phenomenological description of the coupled SUSY theories. Guided by the theory results, in Sec.~\ref{sec:num} we present a full-fledged numerical analysis of the phase diagram. Finally, we draw our conclusions and summarize the outlook in Sec.~\ref{sec:conclu}. 

\section{Model Hamiltonian and overview of the phase diagram}\label{sec:ham}

We study a square ladder of spinless hard-core bosons at $\nu=2/5$ filling (on average two filled sites for every five sites in each chain). The system Hamiltonian is schematically depicted in Fig.~\ref{fig:Ryd}\textbf{(a)}, and reads:
\begin{equation}\label{eq:models}
\begin{split}
    H = &-t\sum_{i,\ell} \left(b_{i,\ell}^\dagger b_{i+1,\ell} + H.c.\right) + U\sum_i n_{i,+}n_{i,-}\\
    & + V\sum_{i,\ell}\sum_{j=1}^{r_C}n_{i,\ell}n_{i+j,\ell}.
\end{split}
\end{equation}
where $b_{i,\ell}^\dagger$ is the creation operator for hard-core bosons on site $i$ of chain $\ell=+,-$ and $n_{i,\ell}=b_{i,\ell}^\dagger b_{i,\ell}$. $t$ is the intra-chain hopping, $U$ is the nearest neighbor inter-chain interaction amplitude, $V$ is the longer-ranged intra-chain interaction amplitude, and $r_C \in \mathbb{N}$. The density $\nu=2/5$ is chosen in such a manner that, in the classical limit, the model supports clustering. In this model, the U(1) charges of the two legs are conserved separately, making the Hamiltonian a U(1)$\times$U(1) symmetric system.
There is no inter-chain tunneling but the two legs interact. 

The model in Eq.~\eqref{eq:models} is directly inspired by the Rydberg experiments performed in Ref.~\cite{Henkel2010}. There, a gas of atoms was weakly coupled to Rybderg $p$-state in the Paschen-Bach regime, resulting in pairwise interactions of the type:
\begin{equation}\label{eq:dressing}
    V(r)=V_{\text{max}}/\left(1+r^6/r_S^6\right),
\end{equation}
with both $V_{\text{max}}>0$ and $r_S$ easily tunable in both the transversal and longitudinal directions. Tunneling between tubes was instead strongly suppressed leveraging on the comparatively large quadratic Zeeman shifts from tube to tube. An alternative way to prevent inter-wire tunneling is to suppress it utilizing a higher potential barrier between the wires. Densities could be tuned via changing the atom loading scheme.

The potential in Eq.~\eqref{eq:dressing} is connected to the model Hamiltonian above as follows. In a lattice, when $r_S \sim a_0$ with $a_0$ being the lattice spacing, such potential is well approximated by a Heaviside step function stepping down at $r=r_C a_0=\text{Floor}(r_S/a_0)a_0$. We take $r_C=2$ in Eq.~\eqref{eq:models} along the legs and $r_C=1$ along the rungs. Note that anisotropic potentials could be generated in various ways -- either changing the relative lattice spacing along and perpendicular to the wire, or utilizing dressing via $p$- and $d$-states under specific directions. The influence of longer-range terms (for $V$) is known to affect clustering only at the quantitative level~\cite{Mattioli2013}, so, throughout the work, we will solely employ Eq.~\eqref{eq:models} with $r_C=2$ for the sake of clarity.

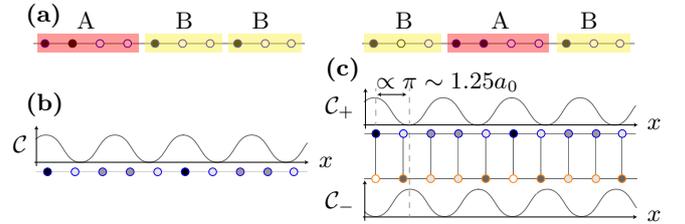
\begin{figure}[t] \begin{overpic}[width=0.97\linewidth]{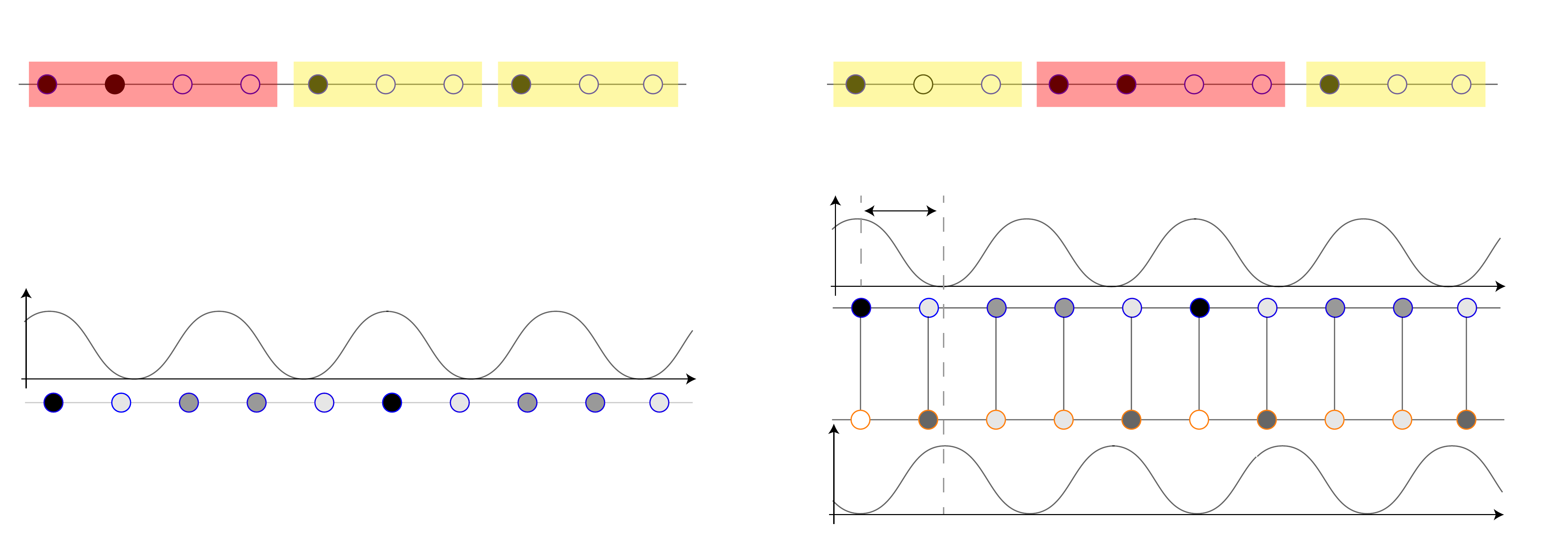}
        \put (0,18) {{\textbf{(b)}}} \put (47,23.5) {{\textbf{(c)}}}  \put (8,31) {{A}} \put (23.5,31) {{B}} \put (35.5,31) {{B}} \put (58,31) {{B}} \put (73,31) {{A}} \put (87,31) {{B}} \put (0,32) {{\textbf{(a)}}} \put (-2,11.4) {{$\mathcal{C}$}} \put (46,9) {{$x$}} \put (47,17.5) {{$\mathcal{C}_+$}} \put (47,2.5) {{$\mathcal{C}_-$}} \put (97.5,15) {{$x$}} \put (97.5,0.5) {{$x$}} \put (55,21.5) {{$\propto \pi \sim 1.25 a_0$}}
    \end{overpic}
    \caption{(Color online.) Cartoon pictures of the leading instability of the various phases in the single chain and ladder model (for $2/5$ filling). \textbf{(a)} Two of the possible semi-classical configurations of the two clusters of hard core bosons A and B in a ratio 1:2. \textbf{(b)} TLL for only one chain. $\mathcal{C}$ is the density-density correlations without the algebraic damping. \textbf{(c)} Spin density wave along $z$-direction for the square lattice for a gapped spin sector. Such a wave propagates with a momentum of $\pm 2k_F$.}
    \label{fig:clusters}
\end{figure}

The red line ($U=0$) of Fig.~\ref{fig:Ryd}\textbf{(b)} corresponds to two decoupled chains of known behavior~\cite{Mattioli2013,Dalmonte2015}. The phase diagram of a single chain displays a Tomonaga-Luttinger liquid (TLL) for $0 < V/t \lesssim 5.7$, and a cluster Luttinger liquid (CLL) for $V/t \gtrsim 5.7$. A CLL behaves like a regular TLL, with its elementary excitations behaving as propagating waves of local (over a few sites) \textit{clusters} of filled and empty sites. For $\nu=2/5$ such clusters minimizes the energy cost in $V$. Two semi-classical cluster configurations are represented in Fig.~\ref{fig:clusters}\textbf{(a)} and involve two clusters A and B in a ratio 1:2. 
Both the TLL and the CLL have a central charge $c=1$. They are separated by a SUSY conformal phase transition point of central charge $c=3/2$ at $V/t \simeq 5.7$ identified numerically~\cite{Dalmonte2015}. The low-energy field theory for this point is described by a combination of a compactified boson and a real Majorana fermion.

By adding an interaction between two of these chains, we infer the rest of the phase diagram Fig.~\ref{fig:Ryd}\textbf{(b)}. We map the leg index to a SU(2) spin-$1/2$ degree of freedom. We show that a partial spin gap opens as soon as $U>0$ (and $V>0$). At weak coupling ($V/t \lesssim 5.7$) the leading instability is a spin-density wave (SDW). At strong coupling ($V/t \gtrsim 5.7$), it is the CLL that is spin-locked. In both cases, the central charge is $c=1$. When $V=0$, the ladder is equivalent to the fully gapless Hubbard model described by a TLL with $c=2$ (2TLL). At $t<U\lesssim V$, we find a charge-density Wave (CDW) leading instability (with $c=1$) between the SDW and the CLL. We understand the existence of this instability here as the consequence of the inversion of the sign of the mass of the spin degree of freedom before clusterization when starting from the SDW and increasing both $U$ and $V$. We provide a cartoon picture of the dominant instabilities in Fig.~\ref{fig:clusters}.

The SDW and CDW phases are separated by the Gaussian phase transition with $c=2$. When $U=0$, the phase transition point at $V/t \simeq 5.7$ is the superposition of the two SUSY conformal phase transition of each independent chain, hence $c=3/2+3/2=3$. The CDW and the CLL are separated by the generalization of the supersymetric phase transition point existing at ($V/t \simeq 5.7$, $U=0$). This transition has $c=3/2$ and may not be supersymmetric. The SDW and 2TLL phases are separated by a Berezinskii-Kosterlitz-Thouless (BKT) transition (green dashed line in Fig.~\ref{fig:Ryd}\textbf{(b)}). Note that the location of this BKT transition is predicted from the analytical results at weak coupling regime and hard to locate precisely for the entire parameter range using numerical tools used in this study.

We establish these results analytically in Sec.~\ref{sec:theory} starting from the two independent chain coupled by the perturbative interaction $U$. We use Abelian bosonization in the weak coupling regime or provide a phenomenological explanation using cluster bosonization for $V/t \gg 5.7$. Moreover, we predict possible outcomes for
the coupling between two SUSY conformal field theories close to ($V/t \simeq 5.7$, $U \gtrsim 0$) using a 
phenomenological field theoretic approach  in Sec.~\ref{sec:effectiveTheory}.
 We use exact diagonalization and infinite density matrix renormalization group (iDMRG) method
in Sec.~\ref{sec:num} to corroborate our theoretical results and predictions, and to uncover the phase diagram of Fig.~\ref{fig:Ryd}\textbf{(b)} for large $U/t$. 

We note that Ref.~\cite{Botzung2019} studied a similar model as Eq.~\eqref{eq:models} but with an inter-leg interactions ($U$) of range two (in diagonal) instead of zero like we do. This corresponds to a very different classical limit, and reflects onto the fact that the phase diagrams, apart from weak coupling regions, are generically distinct. In particular, there appears to be no similar scenario close to the SUSY point. 

\section{Analytical \& phenomenological approach of weakly coupled legs}
\label{sec:theory}

In this section, we derive our analytical results on the ladder model by starting from the two decoupled chains that are either describe by massless single-particles ($V \ll t$) or cluster liquids ($V\gg t$), to which we add the interaction $U$ perturbatively. In the weak coupling regime ($U,V \ll t$), we use Abelian bosonization to predict the SDW phase, the 2TLL phase, and the BKT phase transition in between (the dashed line in Fig.~\ref{fig:Ryd}\textbf{(b)}). In the strong $V$ regime, we use cluster bosonization to describe the properties of two interacting CLL, and how the interaction induces a partial gap, leading to $c=1$ CLL in the ladder geometry. 

\subsection{Weak intra-chain interactions: Abelian bosonization}
\label{sec:weak_coupl_square}

In the weak coupling, we predict the existence of two phases: the 2 Tomonaga-Luttinger liquid when $V=0$ and the holonic phase with spin density wave leading instability when $V>0$ (c.f. Figs.~\ref{fig:clusters}\textbf{(b)} and \textbf{(c)} for a cartoon picture). Formulated as a soft-shoulder Hubbard model of spinful fermions and using Abelian bosonization, we show that the model in Eq.~\eqref{eq:models} describes a TLL with a mass term for the spin degree of freedom which is relevant as soon as $V>0$ at first loop of renormalization and irrelevant when $V=0$ (for the filling $\nu=2/5$). When $U=0$, the model is gapless for both degrees of freedom. 

We obtain these predictions by bosonizing the Hamiltonian in Eq.~\eqref{eq:models} and obtaining explicit formulae for the parameters of the low-energy field theory. Indeed, when both $U=0$ and $V=0$, Eq.~\eqref{eq:models} describes two chains of free spinless fermions (via, e.g., a pair of Jordan-Wigner transformations) with an emerging conformal symmetry when taking the continuous limit. We therefore study the weak-coupling low-energy properties of the system using standard Abelian bozonisation techniques as detailed in Refs.~\cite{DiFrancesco1997,Giamarchi2003,Gogolin2004}. We use the same conventions as in Ref.~\cite{Gogolin2004}:
\begin{subequations}
\begin{align}
    c_{i,\ell}&\sim \psi_{R,\ell}(x=i) + \psi_{L,\ell}(x=i),\\
    \psi_{r,\ell}&=\frac{1}{\sqrt{2\pi a_0}}U_{r,\ell} e^{i r k_F x}e^{i\sqrt{\frac{\pi}{2}}\left(r\phi_c-\theta_c+\ell\left(r\phi_s-\theta_s\right)\right)}.
\end{align}
\end{subequations}
Here, $c_{i, \ell}$ is the fermionic annihilation operator obtained from $b_{i, \ell}$ after a Jordan-Wigner transformation, $\psi_{r,\ell}$ is the associated right ($r=R$) and left ($r=L$) moving fermionic field in the continuous limit close to the Fermi points, $\ell=+,-$ denotes the chain index, $a_0=1$ is the lattice spacing, $U_{r,\ell}$ are the Klein factors such that $\lbrace U_{r,\ell}, U_{r^\prime,\ell^\prime}\rbrace= 2 \delta_{r,r^\prime} \delta_{\ell,\ell^\prime}$, and $k_F=\pi \nu / a_0$ is the Fermi momentum with $\nu=2/5$ being the filling factor for each chain. The bosonic fields $\phi_c$, $\phi_s$, $\theta_c$, and $\theta_s$ are the charge and spin fields and their respective dual fields.

The Hamiltonian in Eq.~\eqref{eq:models} in terms of the bosonic fields can be written as
\begin{equation}\label{eq:boso}
\begin{multlined}[0.8\linewidth]
    H\sim \frac{u_c}{2}\int \mathrm{d}x \left(K_c :\left(\partial_x \theta_c \right)^2: +\frac{1}{K_c} :\left(\partial_x \phi_c \right)^2:\right) \\
     \qquad + \frac{u_s}{2}\int \mathrm{d}x \left(K_s :\left(\partial_x \theta_s \right)^2: +\frac{1}{K_s} :\left(\partial_x \phi_s \right)^2:\right) \\
     \qquad - \frac{g_\perp}{4\pi^2}\cos(\sqrt{8\pi}\phi_s),
\end{multlined}
\end{equation}
where $\partial_x$ is the first derivative with respect to the spatial coordinate $x$, the colon denotes normal ordering, and $a_0$ is the lattice spacing (taken equal for both the $x$ and $y$ directions for simplicity). The quantities
\begin{subequations}\label{eq:parameters}
\begin{align}
    v_F &= 2t \sin(2\pi/5 ),\\
    u_s K_s &= u_c K_c = v_F,\\
    g_\perp&=-2U,\\
    g_\parallel&=-2(U-5V),\\
        K_c&= \frac{1}{\sqrt{1+\frac{U+5V}{\pi v_F}}}, \label{eq:paramsd}
    \\ 
        K_s&=\frac{1}{\sqrt{1+\frac{g_{\parallel}}{2\pi v_F}}},
\label{eq:paramse}
\end{align}
\end{subequations}
are the Fermi velocity $v_F$, bare charge velocity $u_c$, charge Luttinger parameter $K_c$ at first order in $U/t$ and $V/t$, spin velocity $u_s$ and spin Luttinger parameter $K_s$ at first order. Expressions for generic $k_F$ and $r_C$ can be found in the Appendix. The renormalization equations at first loop of renormalization are
\begin{subequations}\label{eq:RG}
\begin{align}
    \frac{dg_{\perp}}{dl} &= \frac{1}{2\pi}g_{\perp}g_{\parallel} \\ \frac{dg_{\parallel}}{dl} &= \frac{1}{2\pi}g^2_{\perp},
\end{align}
\end{subequations}
which predict a BKT phase transition when $g_\parallel=-\lvert g_\perp \rvert$, i.e., when $V=0$ for $U>0$~\cite{Gogolin2004}. 

Solving Eqs.~\eqref{eq:RG} shows that the spin sector is gapped when $V>0$, and gapless when $V=0$~\cite{Gogolin2004}. As the extended interaction breaks the SU(2) symmetry, the renormalization flow may introduce a gap in the spin sector without commensurability effect. As $m = - g_\perp / (4\pi^2) >0$, this gap fixes $\phi_s=\sqrt{\pi/8}$ (modulo $\sqrt{\pi/2}$) such that the order parameter $\mathcal{O}_{\mathrm{triplet}_0}$ is the ``triplet 0''~\cite{Nakamura2000}:
\begin{equation}\label{eq:triplet0}
    \mathcal{O}_{\mathrm{triplet}_0}= \sin(\sqrt{2 \pi}\phi_s).
\end{equation}
This order parameter describes the spin sector contribution in the ``$2k_F$'' spin density wave along the $z$-direction (SDW$^z$) whose order parameter $\mathcal{O}^{z}_{\mathrm{SDW}} $~\footnote{
The contribution also appears in the bond-spin density wave along the $z$-direction (BSDW$^z$) order parameter. As the charge sector is gapless, SDW$^z$ and BSDW$^z$ leading instabilities are indistinguishable.} can be defined as follows~\cite{Dziurzik2006}:
\begin{equation}
\begin{split}
    \mathcal{O}^{z}_{\mathrm{SDW}}&=S^z (x) - \frac{1}{\sqrt{2\pi}}\partial_x \phi_s (x)\\
    &=\cos(\sqrt{2\pi}\phi_c+2k_F x)\sin(\sqrt{2\pi}\phi_s).
\end{split}
\end{equation}
where $S^z (x)$ the continuous limit of the spin operator
\begin{equation}
  \frac{1}{2}  \sum_{\ell, \ell' = \pm}b^\dagger_{x,\ell}\sigma^z_{\ell \ell'}b_{x,\ell'}\sim S^z (x),
\end{equation}
with $\sigma^z_{\ell \ell'}$ the Pauli matrix along the $z$-direction.
The associated correlator is
\begin{equation}\label{eq:SDW_spin}
    \langle \mathcal{O}^z_{SDW}(x)\mathcal{O}^z_{SDW}(y) \rangle \sim \cos(2k_F|x-y|)|x-y|^{-K_c},
\end{equation}
which is also the correlator with the longest range, such that the SDW$^z$ is the dominant instability in the system. The cartoon picture of the low energy states of the phase at low filling is a set of right or left moving particles of momentum close to $2k_F$ distributed equally between the two legs. When two particles from different legs scatter, they exchange momentum.

When $V=0$, the phase is fully gapless as in the Hubbard model. Both the CDWs with momentum $2k_F$ and SDWs instabilities are in competition. We define the order parameter associated with CDW with
\begin{equation}
    \mathcal{O}_{\mathrm{CDW}}=\rho(x) - \sqrt{\frac{2}{\pi}}\partial_x \phi_c (x),
\end{equation}
where $\rho(x)$ is the continuous limit of the density operator
\begin{equation}
    \sum_{\ell=\pm} b^\dagger_{x,\ell} b_{x,\ell}\sim \rho (x).
\end{equation}
The correlation functions associated with both CDW and SDW$^z$ are
\begin{subequations}\label{eq:CDW_both}
\begin{align}
    \begin{multlined}[t][.80\linewidth]
        \langle  \mathcal{O}_{\mathrm{CDW}_{2k_F}}(x)\mathcal{O}_{\mathrm{CDW}_{2k_F}}(y) \rangle \sim  \\ \cos(2k_F|x-y|)|x-y|^{-(K_c+K^*_s)}, 
    \end{multlined}\\
    \begin{multlined}[t][.80\linewidth]
        \langle \mathcal{O}^z_{SDW}(x)\mathcal{O}^z_{SDW}(y) \rangle \sim \\ \cos(2k_F|x-y|)|x-y|^{-(K_c+K^*_s)},
    \end{multlined}
\end{align}
\end{subequations}
with $K^*_s=1$, the renormalized Luttinger parameter as the SU(2) symmetry emerges asymptotically. As a comparison, the CDW correlation function in the SDW phase decays exponentially as
\begin{align}
    \begin{multlined}[t][.80\linewidth]\label{eq:SDW_charge}
        \langle  \mathcal{O}_{\mathrm{CDW}}(x)\mathcal{O}_{\mathrm{CDW}}(y) \rangle \sim \\ \cos(2k_F|x-y|)|x-y|^{-K_c} e^{-|x-y|/\xi}, 
    \end{multlined}
\end{align}
where $\xi \sim \frac{1}{M}$ and $M$ is the dynamically generated mass gap~\cite{Gogolin2004} defined as
\begin{equation}\label{eq:mass-scal}
    M = \Lambda \exp(-\left(g^2_{\perp}-g^2_{\parallel}\right)^{-\frac{1}{2}}),
\end{equation}
with $\Lambda$ being the ultra-violet cutoff.

When $0<V<U/5$, the partial gap of the SDW is more fragile than when $V>U/5$. Indeed, the scaling dimension of the mass operator in the sine-Gordon model in Eq.~\eqref{eq:boso} is larger than two when $K_s>1$, i.e., when $U=5V$. In this zeroth loop approach, the mass operator is therefore irrelevant when $0<V<U/5$ such that the spin gap opens only because of first-loop corrections. 
This feature of the BKT phase transition hints at the difficulty to precisely pinpoint the phase transition line within the regime $0<V<U/5$ seen in our later numerical approach.

\subsection{Strong intra-chain interactions and the clustering limit: cluster bosonization}
\label{sec:CLL_square}

In the strong coupling regime $V\gg t, U$, we show that the phase is a CLL: to minimize the shoulder potential $V$, the hard-core bosons on each leg group up in clusters separated by $r_C$ empty sites. Both the number of particle per cluster and the variety of clusters depend on the filling and $r_C$. For $\nu=2/5$ and $r_C=2$, two clusters A and B emerges with ratio 1:2 and are represented in Fig.~\ref{fig:clusters}\textbf{(a)}. In the CLL, it is the collective modes of these clusters A and B that propagate freely, similarly to the holons of Sec.~\ref{sec:weak_coupl_square}. We show that the inter-leg interaction $U$ may lead to the equivalent of a spin gap for these clusters, forcing the clusters from both legs to propagate in a correlated manner. When this cluster spin gap is present, the central charge is $c=1$, and $c=2$ otherwise. Numerical simulations conclude $c=1$ in Sec.~\ref{sec:num}.

To derive the results, we use cluster bosonization~\cite{Mattioli2013,Dalmonte2015}. The cluster bosonization is a generalization of the phenomenological bosonization~\cite{Giamarchi2003}, also known as the effective harmonic-fluid approach~\cite{Haldane1981}. The idea of the approach is first to identify the clusters and the associated variables in the bosonic field formalism, and then write the effective low-energy Hamiltonian in terms of these cluster bosonic variables. This Hamiltonian is interpreted similarly to a standard bosonized Hamiltonian like Eq.~\eqref{eq:boso}. We define both a cluster charge and a cluster spin degrees of freedom by analogy.

To obtain the cluster bosonic fields, we first reformulate the density distribution after coarse graining the system into the cluster A and B. The corresponding cluster density distribution for each leg ($\ell = \pm$) is
\begin{equation}
    \rho_{\ell}(x) = \sum^{M}_{m=1}f_{\ell}(x_m)\delta(x-x_m),
\end{equation}
where $M$ is the number of clusters on each leg, $f_{\ell}(x_m)=\{1,2\}_{\ell}$ is the coarse graining weight for 1- and 2-particle clusters (B and A clusters, respectively), and $x_m$ is the position of a cluster. By construction, $\sum_{x_m}f_{\ell}(x_m)=N$ where $N$ is the conserved total number of particles. We define the cluster field $\phi_{\ell}(x)$ accounting for the quantum fluctuations of the cluster density. By construction, the cluster field is strictly monotonic and an integer multiple of $\pi$ at the center of a cluster: $\phi_{\ell}(x_m) = \pi m$ with $m \in \mathbb{Z}$, such that
\begin{equation}\label{eq:rho_phi}
    \rho_{\ell}(x) = \nabla\phi_{\ell}(x) \sum^{M}_{m=1}f_{\ell}(x_m)\delta(\phi_{\ell}(x)-\pi m).
\end{equation}
Eq.~\eqref{eq:rho_phi} is defined for $0\leq x \leq L$ where $L$ is the length of the system. For periodic boundary condition, we extend Eq.~\eqref{eq:rho_phi} over all space. We set $\phi_{\ell}(x+L) = \phi_{\ell}(x) + M\pi$, 
such that for $j=nM+m$ with $1\leq m \leq M$ and $n\in \mathbb{Z}$, we get $x_j=x_m$ modulo $L$,
and we take $\Sh_{\pi M}$ as the unit Dirac comb of periodicity $\pi M$. Then the $L$-periodic Eq.~\eqref{eq:rho_phi} is 
\begin{subequations}\label{eq:rho_phi_2}
\begin{align}
    \rho_{\ell}(x) &= \nabla\phi_{\ell}(x) \sum^{+\infty}_{j=-\infty}f_{\ell}(x_j)\Sh_{\pi M}(\phi_{\ell}(x)-\pi j), \\
    &=\frac{N}{M} \nabla\phi_{\ell}(x) \sum^{+\infty}_{q=-\infty}\frac{A_{q,\ell}}{\pi}e^{-i2q\phi_{\ell}(x)},
\end{align}
\end{subequations}
where we have used Poisson's summation, and defined $NA_{q, \ell}$, $q\in \mathbb{Z}$, as the Fourier coefficients of $f_{\ell}(x_j)$, a function of $j$ of period $M$. The coefficients $A_{q,\ell}=A^*_{-q,\ell}$ are not universal and depend on the microscopic details of the model.

In the density-angle variables formalism, we define the amplitude of the cluster bosonic field as the square-root of the coarse grained (or cluster) density, and the cluster angle is the associated canonical conjugate variable. To construct these fields, we introduce $\phi'_{\ell}(x)$~\footnote{We used the factor $\sqrt{\pi}$ in Eq.~\eqref{eq:phiprime} to obtain an expression for the bosonized Hamiltonian directly comparable to Sec.~\ref{sec:weak_coupl_square}.} as 
\begin{equation}\label{eq:phiprime}
    \phi_{\ell}(x) = \pi \nu \sigma x - \sqrt{\pi}\phi'_{\ell}(x),
\end{equation}
where $\sigma = M/N = 3/4$ and $\nu = N/L= 2/5$. The cluster density operator is
\begin{equation}\label{density-cbosonization}
\begin{split}
     \rho_{\ell}(x) &= \left(\frac{2}{5} - \frac{1}{\sqrt{\pi}\sigma}\nabla\phi'_{\ell}(x)\right)\\
     & \qquad \times \sum^{+\infty}_{q=-\infty}A_{q,\ell}e^{-i2q(3\pi x/10 - \sqrt{\pi}\phi'_{\ell}(x))}.
\end{split}
\end{equation}
We define $\theta'_{\ell}(x)$ as the conjuguate variable of $\nabla \phi'_{\ell}(x)$ such that
\begin{equation}
    \left[\theta'_{\ell}(y),\frac{1}{\pi}\nabla \phi'_{\ell'}(x)\right] = i\delta_{\ell,\ell'}\delta(x-y).
\end{equation}
We deduce the expression for the cluster bosonic field $\psi^{\ph}_{\ell}(x)$ in the density-angle formulation
\begin{equation}\label{field-cbosonization}
\begin{split}
    \psi^{\ph}_{\ell}(x) =& e^{-\frac{i\sqrt{\pi}}{\sigma}\theta'_{\ell}(x)}\sqrt{\frac{2}{5} - \frac{1}{\sqrt{\pi}\sigma}\nabla\phi'_{\ell}(x)}
    \\
    & \qquad\sum^{+\infty}_{q=-\infty}\alpha_{q,\ell}e^{-i2q(3\pi x/10 - \sqrt{\pi}\phi'_{\ell}(x))},
\end{split}
\end{equation}
where $\alpha_{q,\ell}$ are the non-universal Fourier coefficients of the square root of the sum in Eq.~\eqref{density-cbosonization}.

Using the cluster density (Eq.~\eqref{density-cbosonization}) and the cluster bosonic field (Eq.~\eqref{field-cbosonization}), we write the effective continuous Hamiltonian describing the low-energy behavior of the clusters. For $U=0$, this Hamiltonian is~\cite{Mattioli2013,Dalmonte2015}
\begin{equation}
\label{clusterv=0}
H = \sum_{\ell = \pm}\frac{v_{\ell}}{2}\int dx\left(K_{\ell}(\nabla \theta'_{\ell})^2 +K^{-1}_{\ell}(\nabla \phi'_{\ell})^2 \right),
\end{equation}
where the cluster velocities and Luttinger parameters include the (unknown) Jacobian of the coarse graining. By symmetry both $v_{\ell}=v$ and $K_{\ell}=K$ for the two chains. The Hamiltonian in Eq.~\eqref{clusterv=0} describes two CLL (2CLL), with central charge $c=1$ each ($c=2$ in total)~\cite{Mattioli2013,Dalmonte2015}. The inter-leg interaction in the cluster bosonic field variables is
\begin{equation}
\label{clusterhubbard}
\begin{split}
    U \int dx \rho_{+}(x)\rho_{-}(x) \sim & \ g_0\int dx \nabla \phi'_{+}\nabla \phi'_{-}  \\
     + & g_\perp\int dx \cos(\sqrt{4\pi}(\phi'_{+}-\phi'_{-}))
    \\
    + & g_2\int dx \sin(\sqrt{4\pi}(\phi'_{+}-\phi'_{-}))+...
\end{split}
\end{equation}
with $g_0=\frac{UA_{0,+}A_{0,-}}{\pi\sigma^2}$, $g_\perp=8U\operatorname{Re}[A_{1,+}A_{-1,-}]/25$ and $g_2=-8U\operatorname{Im}[A_{1,+}A_{-1,-}]/25$. In Eq.~\eqref{clusterhubbard}, we have only written the slowly oscillating terms at least as relevant as $\cos(\sqrt{4\pi}\phi^\prime)$. By symmetry, at zeroth order in perturbation of $U$, the coefficient $A_{q,\ell}$ are independent of the leg index. In this case, we expect $g_0>0$, $g_\perp>0$ and $g_2=0$.  We define the cluster spin and charge bosonic field $\phi_{s}$ and $\phi_{c}$ as
\begin{subequations}\label{spin-charge}
\begin{align}
&\phi_{c} = \frac{\phi_{+} +  \phi_{-}}{\sqrt{2}}, &\phi_{s} = \frac{\phi_{+} -  \phi_{-}}{\sqrt{2}},\\
&\theta_{c} = \frac{\theta_{+} + \theta_{-}}{\sqrt{2}}, & \theta_{s} = \frac{\theta_{+} - \theta_{-}}{\sqrt{2}}.
\end{align}
\end{subequations}
Using Eqs.~\eqref{clusterv=0}-\eqref{spin-charge}, we derive the Hamiltonian in Eq.~\eqref{eq:models} with inter-leg interaction as follows:
\begin{equation}\label{eq:square_cluster}
\begin{multlined}[0.8\linewidth]
    H \sim \frac{v_c}{2}\int\left(K_c(\nabla \theta'_c)^2 + K^{-1}_c(\nabla \phi'_c)^2 \right) 
    \\
    +\frac{v_s}{2}\int\left(K_s(\nabla \theta'_s)^2 + K^{-1}_s(\nabla \phi'_s)^2 \right)
    \\
    +g_\perp\int dx \cos(\sqrt{8\pi}\phi'_{s}) ,
\end{multlined}
\end{equation}
 where
\begin{subequations}\label{eq:Cparameters_square}
\begin{align}
    K_c &= \frac{K}{\sqrt{1+\frac{g_0K}{v}}}, \\
    K_s &= \frac{K}{\sqrt{1-\frac{g_0K}{v}}},
    \label{eq:Cparameters_squareb}
    \\
    v_c&=v\sqrt{1+\frac{g_0K}{v}},\\
    v_s&=v\sqrt{1-\frac{g_0K}{v}}.
\end{align}
\end{subequations}

According to this phenomenological picture, we expect the cluster charge sector to be gapless, while the spin sector may be gapped or gapless. The partial gap depends on the value of $K_s$, and leads to a CLL with $c=1$ or $c=2$. As the inter-leg interaction ($U$) does not involve the cluster charge bosonic field, Eq.~\eqref{eq:square_cluster} displays spin-charge separation with is no  gap opening when $U=0$~\cite{Mattioli2013,Dalmonte2015}. 
When $K_s>1$, the cluster spin mass is irrelevant. In both of these cases, both the charge and the spin sectors are gapless such that the system describes a 2CLL with $c=2$ at zero loop of renormalization. When $K_s<1$, the cluster spin gap is relevant. Like for the SDW of Sec.~\ref{sec:weak_coupl_square} we expect the mass to be positive.
Unlike the SDW, the coarse grained nature of the cluster degree of freedom prevents an easy interpretation of the microscopic correlation functions obtained numerically. Given the repulsive nature of $U$, we nonetheless can expect an anti-alignment of the clusters along the rung. Then the phase is a spin-locked 1CLL with $c=1$. The numerical simulations of Sec.~\ref{sec:num} differentiates the 2CLL from the 1CLL, and show that the phase is indeed the spin-locked CLL with $c=1$ as soon as $U>0$.

\section{Coupling SUSY conformal field theories: a phenomenological approach}
\label{sec:effectiveTheory}

In the intermediate coupling regime that captures the transition between weak and strong coupling, none of the above approaches is reliable. In particular, close to the transition point at $U=0$, $V/t\simeq 5.7$, each chain is separately described by a $c=3/2$ supersymmetric conformal field theory, composed of a compactified boson and a real fermion field with the same speed of sound. Coupling two chains thus presents a very rare opportunity to investigate the coupling of supersymmetric theories  -- something that, to the best of our knowledge, is not achievable in other cold gas settings. 

In order to provide a qualitative understanding of the system dynamics in the vicinity of the SUSY point, we first recap what is known about the single chain, and then develop an effective, phenomenological field theory for the ladder. The latter theory results in two possible scenarios for the phase diagram. We resolve such dichotomy by means of tensor network simulations in the next section.  It is important to note that we are interested here in the intermediate-interaction parameter regime, so that our approach is complementary to hard-constrained models~\cite{Bauer2013}.

\subsection{Brief recap: SUSY critical point at $U=0$}

For a single chain, the vicinity of the SUSY phase transition point at $V/t\simeq 5.7$ can be phenomenologically described by the field theory of a massless compact bosonic field with $c=1$ and a real fermion (with $c=1/2$ when massless). Specifically, the effective Hamiltonian around the SUSY conformal point reads
\begin{equation}\label{eq:susy_1D}
\begin{split}
    H^{\text{1SCF}} &= \frac{v_{\text{B}}}{2}\int dx \left( (\partial_x \varphi)^2 + (\partial_x \vartheta)^2\right),
    \\
    & \qquad +i\int dx \left(v_{M}\eta\partial_x\zeta + m_M \eta\zeta\right),
\end{split}
\end{equation}
where $\varphi$ and $\vartheta$ are the two conjugate compact bosonic fields of associated velocity $v_B$. The fields are not the microscopic bosonic field nor the cluster fields: in fact, their relation to the microscopic operators has not been determined. Over the transition regime, they do not acquire a mass. $\eta$ and $\zeta$ are a pair of Majorana operators of velocity $v_{M}$ and mass $m_M$. At the SUSY conformal point
$m_M=0$: at that point, $v_F=v_B$, and the Luttinger parameter of the bosonic field is fixed by the SUSY~\cite{Bauer2013}. 

Based on a level spectroscopy analysis, one can interpret the phase transition towards the CLL as an inversion of the sign of $m_M$~\cite{Dalmonte2015}. At the phenomenological level, one can interpret the real fermion as an Ising field, that is `ordered' in the CLL phase, and `disordered' in the TLL phase. It is thus a field that indicates the presence of composite particles as fundamental objects. It is worth mentioning that the appearance of a critical point with central charge larger than one, while certainly unusual for a single-species model, is not fully unexpected: indeed, it is known that by considering quantum dynamics beyond nearest-neighbor, even free theories can display phases with more than one-gapless channel - one paradigmatic example being tight-binding models where NN and NNN tunnelings are of the same order~\footnote{We thank M. Fabrizio for pointing out this connection to us.}.

\subsection{Coupling two SUSY critical points}

Starting from the above picture, we develop a description to the ladder scenario. In the ladder, there are two compact bosonic degrees of freedom: their charge combination acts as an underlying field that remains gapless over the transition regime, while the spin combination shall be gapped to account for the presence of the SDW. 
The real fermion theory is also doubled. 

Based on these two observations, the effective Hamiltonian in the vicinity of the SUSY conformal phase transition point $U=0,V/t\simeq 5.7$ reads:
\begin{equation}\label{eq:susy_2D}
\begin{split}
    H^{\text{2SCF}} &= \frac{v_{\text{B,c}}}{2}\int dx \left( (\partial_x \varphi_c)^2 + (\partial_x \vartheta_c)^2\right),
    \\
    & \qquad + \frac{v_{B,s}}{2} \int dx ((\partial_x \varphi_{s})^2 + (\partial_x\vartheta_s)^2) \\
    & \qquad + g\int dx\cos(\sqrt{8\pi K_s}\varphi_s)\\
    & \qquad +i\int dx \left(v_{M,S}\eta_S\partial_x\zeta_S + m_{M,S} \eta_S\zeta_S\right)\\
    & \qquad +i\int dx \left(v_{M,A}\eta_A\partial_x\zeta_A + m_{M,A} \eta_A\zeta_A\right),
\end{split}
\end{equation}
where the indices $c$, $S$, $s$, and $A$ stand for charge, symmetric, spin and anti-symmetric respectively. The charge (resp. spin) boson and symmetric (resp. anti-symmetric) Majorana fermions are symmetric (resp. anti-symmetric) combinations of the fields of the two legs. By consistency with our results in Sec.~\ref{sec:weak_coupl_square}, $g>0$ in the SDW phase and the sine-Gordon interaction is relevant. There is no Umklapp term here as neither the weak or strong coupling regimes support an insulting phase.

In principle, other terms could be included in the low-energy theory. Those could either couple the bosonic fields to the fermions, or the two fermions. The latter are likely not relevant to our description: since the system is symmetric under chain inversion, a coupling between anti-symmetric and symmetric clustering real fermionic fields is unlikely. It is instead hard to justify, in general, the absence of terms coupling fermions to bosons. In fact, with the possible exception at the vicinity of the $U=0$ critical point, there is no microscopic reason that suggests not to include them: however, the lack of an exact lattice-to-field operator mapping prevents us from having a clear idea on their possible functional form. We will thus proceed under the assumption that those terms are irrelevant, and verify this a posteriori in the next section.

Under these assumptions, we can treat the bosonic and fermionic sectors of the theory in a modular manner:
\begin{itemize}
    \item charge field $\varphi_c$: this field is always gapless, and describes collective density fluctuations (either starting from a single particle description, or from a cluster one);
    \item spin field $\varphi_s$: this field is strongly pinned in the SDW phase, owing to the value of the spin-Luttinger parameter $K_s\ll 1$. In such regime of the sine-Gordon model, it is still possible for the system to undergo a Gaussian transition at $g=0$: this transition is described by a $c=1$ CFT, and separates two phases with opposite pinning of the spin field;
    \item symmetric fermionic field: we expect this field to undergo an Ising transition, separating a regime where microscopic degrees of freedom are clusters, from the one where those are single particle fields;
    \item anti-symmetric fermionic field: this field might lead to interesting dynamics in case of unbalanced chains (for instance, including clusterization only in one of them). However, since such imbalance is not present in the strong coupling limit if the U(1)$\times$U(1) symmetry is preserved, we do not expect this sector of the theory to be of relevance here.
\end{itemize}

The aforementioned considerations open two possibilities for the quantum criticality. We illustrate these two scenarios moving from weak to strong coupling by, e.g., increasing $V$ and $U>0$, in the following.
\begin{enumerate}[label=P.\arabic*]
    \item In the first one, the spin field undergoes a Gaussian transition first into a CDW phase, and then clustering occurs via an Ising transition. The sequence of central charges here is $c=1$ (SDW), $c=2$ (Gaussian critical line), $c=1$ (CDW), $c=3/2$ (Ising critical line), $c=1$ (spin-locked CLL). \label{pos:1}
    \item The second scenario would instead see first cluster forming, and subsequently, a Gaussian transition in the spin sector.  The sequence of central charges here is $c=1$ (SDW), $c=3/2$ (Ising critical line), $c=1$ (cluster liquid that we cannot fully characterize), $c=2$ (Gaussian transition), $c=1$ (spin-locked CLL). \label{pos:2}
\end{enumerate}
  Here, we rule out the scenario where both transitions happen along the same line, as this will require fine tuning (to the best of our understanding). We note that, while hard to diagnose, it may be possible under both scenarios to recover SUSY along the critical $c=3/2$ line.

Since we are lacking even a phenomenological functional form for the effective parameters in Eq.~\eqref{eq:susy_2D} as a function of $(U,V)$, it is not possible to determine which of the two scenarios of SUSY-breaking is actually taking place. We will thus scrutinize our prediction against numerical simulations in the next section.

\section{Numerical results}\label{sec:num}

To validate the analytical approaches and extend the prediction to larger $U$, we now use exact diagonalization (ED) for small system-sizes and tensor-network (TN) numerical simulations to tackle larger systems sizes. ED (Sec. \ref{subsec:ED}) hints at the phase diagram Fig.~\ref{fig:Ryd}\textbf{(b)}. The picture is completed Sec.~\ref{subsec:iDMRG} by the infinite Density Matrix Renormalization Group method (iDMRG), a variational methods to obtain a matrix-product state (MPS) approximation of the ground state directly in the thermodynamic limit. Using iDMRG, we compute the central charges and useful correlation functions of all phases and transitions.

\subsection{Exact diagonalization}
\label{subsec:ED}

The first step in the numerical analysis of the considered model relies on the exact diagonalization (ED) of the quantum Hamiltonian for small systems-sizes. This method will give access to the physics of the whole phase diagram with accuracy limited only by the finite size of the system \cite{Sandvik2010}.

Since the model has an internal U(1)$\times$U(1) symmetry, we fix the symmetry sector in which we want to analyze the ground state, i.e., the constant filling $\nu_{\ell}$, $\ell = \pm$, for each chain. This restricts us to diagonalize the Hamiltonian for system-sizes such that the total particle number in each chain $N_{\ell} = \nu_{\ell} L$ is an integer.
Furthermore, since we want to understand the effect of interactions in the CLL phase, we also need to consider fillings and system sizes which are commensurate with this phenomenon.
The smallest system size that fulfills these constraints is $L = 10$ for $\nu_{\pm} = 2/5$ and $r_C = 2$. Indeed, 10 sites are needed for $U = t = 0$ and $V>0$ to represent the classical ground states on one chain that are combined together in a CLL ground state by second order perturbation theory in $t$ (see Fig~\ref{fig:clusters}\textbf{(a)}).

\begin{figure}[htb]
\centering
\includegraphics[width=\linewidth]{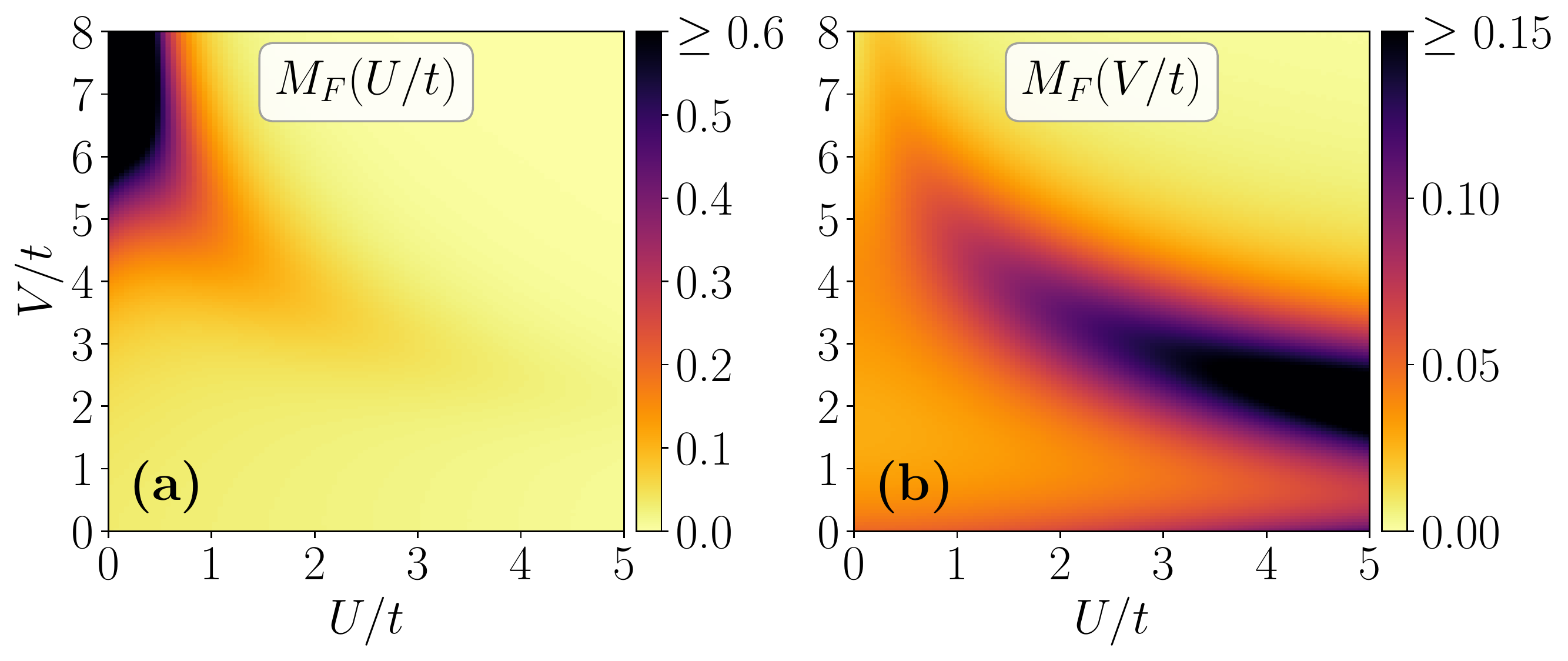}
  \caption{\label{fig:ED_fid} (Color online.) 
  The fidelity-susceptibility $M_F$ (see Eq.~\eqref{eq:fid_sus}) in the $(U/t, V/t)$-plane while varying the parameters \textbf{(a)} $U/t$ and \textbf{(b)} $V/t$.
  Results are obtained via exact diagonalization of a system of size $L=10$ with density $\nu = 2/5$. In \textbf{(a)}, large values of $M_F(U/t)$ near the $U/t =0$ line for $V/t \gtrsim 5.7$ could signal a phase transition between two decoupled CLL (2CLL), that appears for $U/t=0$, to the spin-locked 1CLL phase for $U/t > 0$ (see the text for details).
  On the other hand, in \textbf{(b)}
  a transition between the SDW phase and the CLL phase can be seen. Moreover, for small $V/t$, a weak signature of 2TLL-SDW transition, pertaining to the BKT transition predicted in Sec.~\ref{sec:weak_coupl_square},  is visible.
   }
\end{figure}

An useful tool to verify the overall structure of the phase diagram is provided by the fidelity, defined as 
\begin{equation}
    F(h, \delta h) = |\langle gs(h) | gs(h+\delta h) \rangle|
\end{equation}
with $h$ being either $U/t$ or $V/t$,
as a probe of the phase diagram at different values of $U/t$ and $V/t$. It is known that the fidelity displays a minimum in a finite-size system when crossing what would be a quantum phase transition in the thermodynamic limit~\cite{Zanardi2006}: this is due to the fact that transition lines typically correspond to parameter regimes where the (finite-size) ground state wave function changes very rapidly with respect to changes in the microscopic parameters. At the quantitative level, this is typically better detected by considering the  fidelity-susceptibility~\cite{GU2010}, defined as:
\begin{equation}
    M_F(h) = \lim_{\delta h \rightarrow 0} \frac{-2 \ln F(h, \delta h)}{(\delta h)^2}.
    \label{eq:fid_sus}
\end{equation}
It shows a maximum near the transition (and diverges in the thermodynamic limit at the transition), and is more commonly considered in the literature since it displays universal features at quantum critical points \cite{Schwandt2009}.

The scans of the fidelity-susceptibility $M_F$ in the $(U/t, V/t)$-plane while varying either $U/t$ or $V/t$ are presented in Fig.~\ref{fig:ED_fid}. The fidelity-susceptibility plots suggest the presence of three transitions, which we summarize in the following observations:

\begin{enumerate}
    \item The scan of $M_F(U/t)$ in Fig.~\ref{fig:ED_fid}\textbf{(a)} suggests that for $V/t \gtrsim 5.7$ there is a transition between the decoupled 2CLL phase at $U/t=0$ to an another phase for $U/t > 0$; 
    \item The trends of $M_F(V/t)$ in Fig.~\ref{fig:ED_fid}\textbf{(b)} clearly shows a transition between the SDW and the CLL phase also for $U/t > 0$, as predicted by the theory. At large $U/t>3$, an intermediate phase seems to appear; 
    \item Furthermore, there is a weak signal near small $V/t$ in Fig.~\ref{fig:ED_fid}\textbf{(b)}.
\end{enumerate}

All of the observations above are consistent with the weak coupling theory presented in Sec.~\ref{sec:theory}. In fact, not only the strongest signal confirms the presence of a phase transition between a liquid regime (the SDW phase) and a clustering regime, but there is also a weak signal for the transition from the 2TLL phase to the SDW phase predicted at small $V/t$, and finite $U/t$. The weakness of the signal could be related to the BKT behavior of this transition \cite{Sun_BKT}.

At strong coupling in $V/t$, the simulations  suggest that there might be one transition from 2CLL phase (with $c=2$) to 1CLL phase (with $c=1$) for $U/t \neq 0$ if the cluster spin-gap is relevant when $K_s < 1$ (see Eqs.~\eqref{eq:square_cluster} and \eqref{eq:Cparameters_square}). In the regime of intermediate $V$ and finite $U$, the transition between SDW and CLL seems to split and give rise to an intermediate phase, as predicted by both the scenarios in Sec.~\ref{sec:effectiveTheory}.

However, since we are only restricted to one small system-size of $L=10$ for the ED calculations, the above observations should be interpreted with caution. Moreover, the ED results are unable to draw any conclusions about the two possibilities coming out of the phenomenological model near the SUSY point given in Sec.~\ref{sec:effectiveTheory}, as the horizontal signal in Fig.~\ref{fig:ED_fid}\textbf{(b)} is very broad and can not distinguish between one or two transitions.

\subsection{Tensor network analysis}
\label{subsec:iDMRG}

In this part, we report the results of a numerical analysis using tensor networks (TN) for the system at hand. First we map out the topology of the entire phase diagram of the system in terms of the entanglement entropy, in order to determine the existence of different phases and phase transitions. Specifically, we  show the existence of a charge-density wave (CDW) phase sandwiched between the SDW and the CLL phases. Then we characterize different phase transitions and phases by means of central charges and correlation functions. 

The numerical analysis is based on the widely successful density matrix renormalization group (DMRG) \cite{white_prl_1992, white_prb_1993, white_prb_2005, McCulloch_2007} method using the matrix-product state (MPS) ansatz \cite{schollwock_aop_2011, Orus_aop_2014}. This ansatz relies on the truncation of the Schmidt spectrum, keeping only the $\chi$ (which will be referred to as the bond dimension) largest values, therefore approximating the target state with a quantum state that has area-law entanglement \cite{schollwock_aop_2011}, and
which admits an efficient representation for one-dimensional gapped systems even at large (as well as infinite) sizes.

For our purpose, we probe the system directly at the thermodynamic limit using the infinite DMRG (iDMRG) \cite{McCulloch_2008,Crosswhite_2008} method in translationally invariant infinite MPS (iMPS) representation \cite{Vidal_2007} (see Ref. \cite{Kjall_PRB_2013} for an introduction).
One advantage of considering iDMRG over the standard finite-size DMRG for our purpose is that it allows us to circumvent the strong boundary effects that clustering potentials induce at finite-sizes (see e.g., Refs. \cite{Mattioli2013,Giudici2019}), as they do not occur in iDMRG simulations.

Since our system Hamiltonian enjoys global U(1)$\times$U(1) symmetry corresponding to the conservation of both the densities $\nu_+$ and $\nu_-$, we employ U(1)$\times$U(1) symmetric iMPS ansatz for our simulations \cite{Singh_PRA_2010, Singh_PRB_2011}.
Moreover, as mentioned in the previous subsection \ref{subsec:ED}, to capture the onset of the CLL phase with densities $\nu_{\pm}=2/5$ and $r_C=2$, the iMPS representation needs a unit cell of $L = 10m$ sites with $m \in \mathbb{N}$. In our analysis, we have verified that the results remain unaltered for $m \geq 1$, so that we can faithfully restrict ourselves to the lowest possible unit cell of size $L=10$ sites in the iMPS representation.

For the characterization of different phases and phase transitions of the system, we consider two quantities, namely the system correlation length $\xi$ and the von Neumann entanglement entropy $\mathcal{S}$.
The correlation length $\xi_{\mathcal{O}}$ corresponding to any local operator $\mathcal{O}_j$ is defined by the length scale associated with the correlation function
$\langle \mathcal{O}_j \mathcal{O}_{j+R} \rangle
    - \langle \mathcal{O}_j \rangle \langle \mathcal{O}_{j+R} \rangle \sim \exp \left(-R/\xi_{\mathcal{O}}\right)$.
Then the correlation length $\xi$ of the quantum state is given by the maximum of them as $\xi = \max ( \xi_{\mathcal{O}_1}, \xi_{\mathcal{O}_2}, \xi_{\mathcal{O}_3}, \ldots )$. On the other hand, a given pure quantum state $\ket{\psi}$ belonging to a Hilbert space $\mathcal{H}_{AB} = \mathcal{H}_A \otimes \mathcal{H}_B$ can be written as
$ \ket{\psi} = \sum_{k} \lambda_k \ket{e_k^A} \otimes \ket{e_k^B}$,
where $\lambda_k$'s are the Schmidt coefficients and $|e_k^{(A)B}\rangle$'s form orthonormal basis  in $\mathcal{H}_{A(B)}$. The von Neumann entanglement entropy across the bipartition $A:B$ is then defined as
\begin{equation}
    \mathcal{S} = - \sum_k \lambda_k^2 \ln \lambda_k^2.
\end{equation}
It is to be noted that both the Schmidt coefficients and the entanglement entropy across any bond can be calculated very efficiently in the MPS or iMPS representation~\cite{schollwock_aop_2011}.

The theoretical prediction suggests us that all the possible phases in the system are described by conformal field theories, where both the correlation length $\xi$ and the entanglement entropy $\mathcal{S}$
diverge in the thermodynamic limit. Of course, such divergences cannot be captured by the iMPS ansatz with finite bond dimension $\chi$. Instead, the finite value of $\chi$ will impose a length scale given by a finite value of the correlation length $\xi_{\chi}$ 
\footnote{For an iMPS with finite bond dimension $\chi$, the correlation length $\xi_{\chi}$ can be defined as
\begin{equation*}
    \xi_{\chi} = -1/\ln |\epsilon_2|
\end{equation*}
where $\epsilon_2$ is the second largest eigenvalue of the iMPS transfer matrix \cite{Kjall_PRB_2013}.
} that scales as $\xi_{\chi} \propto \chi^{\kappa}$ with $\kappa$ being a scaling exponent \cite{Tagliacozzo_PRB_2008, Pollmann_PRL_2009, Pirvu_PRB_2012}.
The $\chi$-dependent entanglement entropy $\mathcal{S}_{\chi}$ is then follow the following well-known scaling formula \cite{callan_geometric_1994, vidal_PRL_2003, calabrese_entanglement_2004}:
\begin{equation}
    \mathcal{S}_{\chi} = \frac{c}{6} \ln \xi_{\chi} + b',
    \label{eq:cardy_calabrese_infinite}
\end{equation}
where $c$ is the central charge for the underlying conformal field theory and $b'$ is a non-universal constant.

By varying  the iMPS bond dimension in the range $[64, 1280]$, we use Eq.~\eqref{eq:cardy_calabrese_infinite} to characterize different phases and phase transitions in terms of the central charge $c$. We refine the characterization by computing the correlation functions of the form
\begin{equation}
    \mathcal{C}_{\mathcal{O}}(R) = \langle \mathcal{O}_j \mathcal{O}_{j+R} \rangle
    - \langle \mathcal{O}_j \rangle \langle \mathcal{O}_{j+R} \rangle,
    \label{eq:correlation_function}
\end{equation}
where $\mathcal{O}_j$ is a local operator. When $R > \xi_{\chi}$ all correlations will trivially decay exponentially, restricting the extension of the correlation function to $R \approx \xi_{\chi}$ at maximum.

\subsubsection{The phase diagram: a glimpse from the von Neumann entropy}

\begin{figure*}[thb]
        \includegraphics[width=0.8\linewidth]{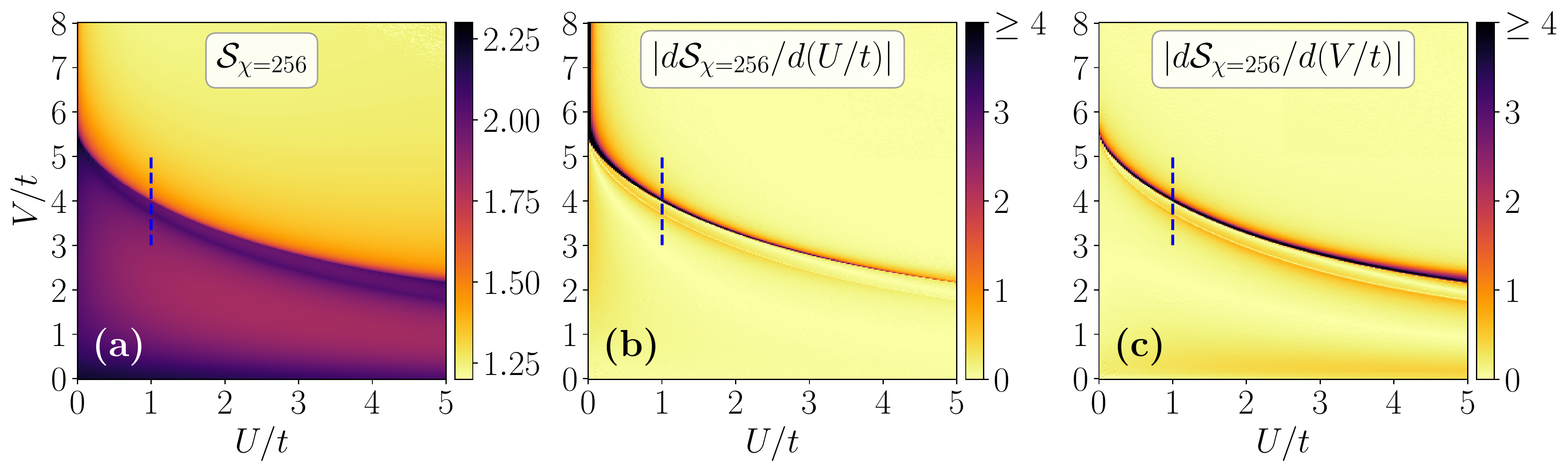}
    \caption{(Color online.) 
    \textbf{(a)} The entanglement entropy $\mathcal{S}$ and  its derivatives with respect to the system parameters \textbf{(b)} $U/t$ and \textbf{(c)} $V/t$  in the $(U/t, V/t)$-plane. The quantities are computed by the iDMRG algorithm with bond dimension $\chi = 256$. The blue dashed lines represent the cut along which the analyses of Fig.~\ref{fig:transition} are performed.
    }
    \label{fig:phase_diagram_num}
\end{figure*}

We present the pattern of the entanglement entropy $\mathcal{S}$ and its derivatives with respect to the system parameters $U/t$ and $V/t$ in the $(U/t, V/t)$-plane in Fig.~\ref{fig:phase_diagram_num} for the iMPS bond dimension of $\chi=256$.
Interestingly, all the scans clearly pick up the signature of the phase transition between the SDW and the CLL phases with increasing $V/t$ for all values of the coupling strength $U/t$. 
Moreover,
similar to the feature of Fig.~\ref{fig:ED_fid}\textbf{(a)}
the derivative of $\mathcal{S}$ with respect to $U/t$ (Fig.~\ref{fig:phase_diagram_num}\textbf{(b)}) also shows the sign of the phase transition between the decoupled 2CLL phase at $U/t=0$ for $V/t \gtrsim 5.7$ to a coupled CLL phase at $U/t > 0$.

However, on a closer inspection of the scans (specifically, Figs.~\ref{fig:phase_diagram_num}\textbf{(a)} and \textbf{(c)}) another signal of a second phase transition at $U/t > 0$ and appearance of another phase sandwiched between the SDW and CLL phases become apparent. This transition and the phase at $U/t > 0$ originates from the $c=3$ SUSY point at $U/t = 0$ and $V/t \simeq 5.7$.
These observations of having another phase transition line and existence of a sandwiched phase corroborate the phenomenological analysis presented in Sec.~\ref{sec:effectiveTheory}.

From the analysis of Fig.~\ref{fig:phase_diagram_num}, it is also clear that the signature of the BKT transition between the 2TLL and the SDW phases at low $V/t$ values cannot be quantitatively captured from such numerical analysis, similarly to what has been observed from the ED results. What we can estimate is a lower bound for the $c=1$ phase based on the study of the central charge, as the latter typically overestimates the extent of critical phases (in our case, the $c=2$ one).

\subsubsection{Coupling SUSY conformal field theories: a numerical perspective}

\begin{figure}
        \includegraphics[width=\linewidth]{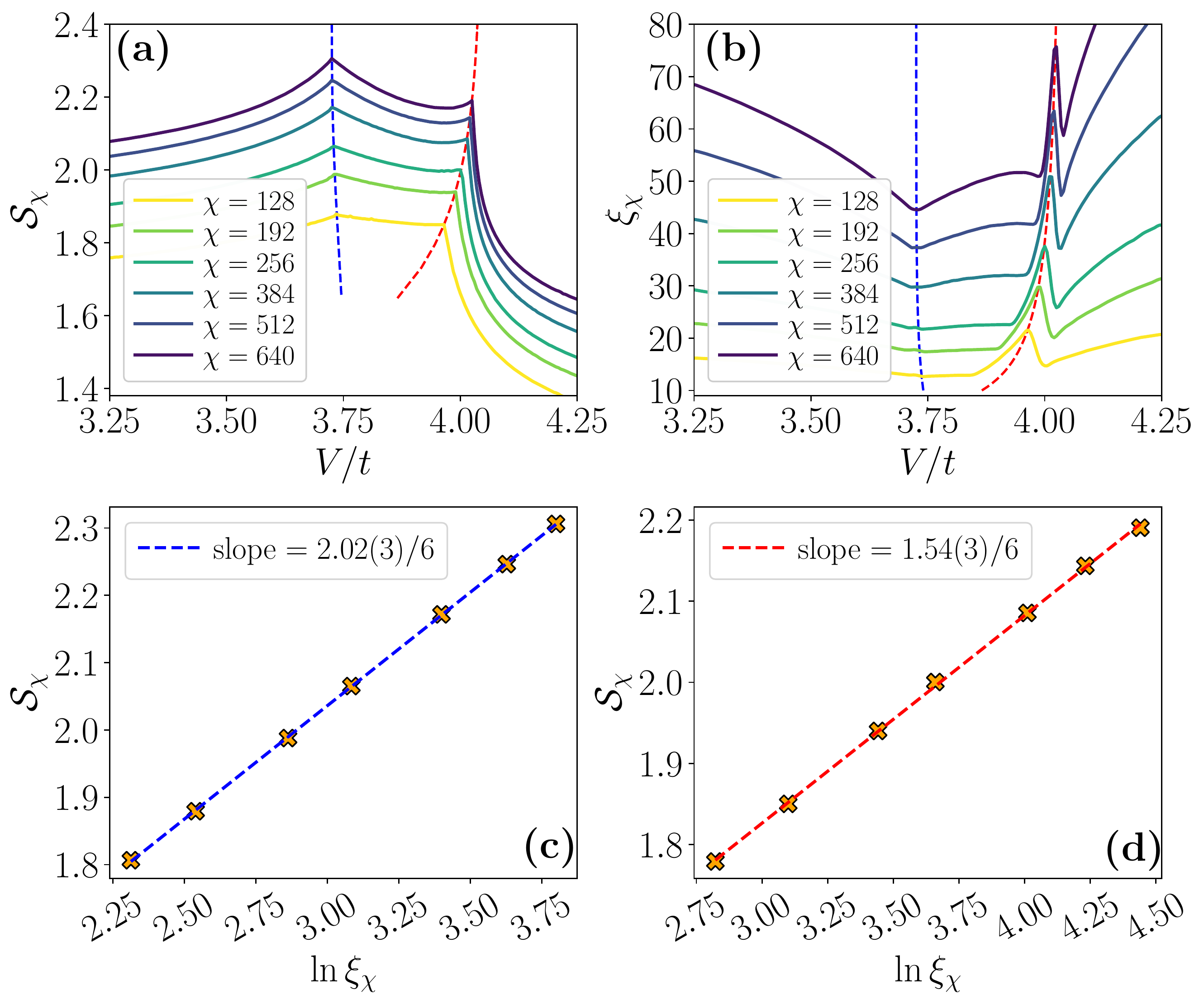}
    \caption{(Color online.) The variations of \textbf{(a)} the entanglement entropy $\mathcal{S}_{\chi}$ and \textbf{(b)} the correlation length $\xi_{\chi}$ with respect to $V/t$ for fixed $U/t=1$ and for different bond dimensions $\chi$. The blue dashed lines in \textbf{(a)} and \textbf{(b)} represent the SDW-CDW transition, while the red dashed lines are for the CDW-CLL transition.
    \textbf{(c)}-\textbf{(d)} The scaling of the entanglement entropy according to Eq.~\eqref{eq:cardy_calabrese_infinite} across the $\chi$-dependent transition points for \textbf{(c)} the SDW-CDW and \textbf{(d)} the CDW-CLL phase transitions. The central charges that we obtain from the scaling for these transitions are respectively $c = 2.02(3)$ and $c = 1.54(3)$.
    }
    \label{fig:transition}
\end{figure}

Based solely on the entanglement entropy analysis, it is already possible to discern whether one of the two proposed scenarios of coupled-SUSY theories presented in Sec.~\ref{sec:effectiveTheory} is describing the vicinity of the $c=3$ critical point.

In Figs.~\ref{fig:transition}\textbf{(a)} and \textbf{(b)}, we present the entanglement entropy $\mathcal{S}_{\chi}$ and the correlation length $\xi_{\chi}$ respectively as functions of $V/t$ for fixed $U/t=1$ with different bond dimensions in the range $\chi \in [128, 640]$.
The variations of both $\mathcal{S}_{\chi}$ and $\xi_{\chi}$ display sharp non-analytic kinks in their profile signalling the presence of two phase transitions -- coherent with both scenarios of Sec.~\ref{sec:effectiveTheory}. The location of the transitions change with respect to the bond dimension $\chi$ (or more precisely with the correlation length $\xi_{\chi}$) following the standard power-law scaling:
\begin{equation}\label{eq:defnu}
    (V_c/t)_{\chi} = (V_c/t)_{\chi \rightarrow \infty} + a \xi_{\chi}^{-\Xi},
\end{equation}
where $V_c$ denotes the transition point and $\Xi$ is a scaling exponent with $a$ being a constant. While $\Xi$ in Eq.~\eqref{eq:defnu} appears to be analogous to the inverse of the thermodynamic critical exponent $\nu$, the two quantities are a priori unrelated. The first is associated with an entanglement-based length scale introduced by the numerical method, whereas the second is associated with a length scale imposed by the inverse of the smallest gap in the system (which is zero at every point of the phase diagram at this filling). In Figs.~\ref{fig:transition}\textbf{(a)} and \textbf{(b)}, the locations of the transitions for different bond dimensions are present by the dashed lines. Importantly, the size of the intermediate phase is sufficiently large so that a finite-size characterization will be possible via correlation functions: we will come back to this point below. 

To precisely obtain the central charges of these transitions, instead of using Eq.~\eqref{eq:cardy_calabrese_infinite} for fixed values of $U/t$ and $V/t$, we perform the same scaling of the entanglement entropy across the $\chi$-dependent transition points (marked by the blue and red dashed lines in Figs.~\ref{fig:transition}\textbf{(a)} and \textbf{(b)}) as prescribed in Ref.~\cite{Kjall_PRB_2013}. Figs.~\ref{fig:transition}\textbf{(c)} and \textbf{(d)} show such scaling across these two phase transitions. Interestingly, the central charges that we obtain from these scaling are compatible at the percent level with $c=2$ for the SDW-CDW transition and $c=3/2$ for the CDW-CLL transition (see Figs.~\ref{fig:transition}\textbf{(c)} and \textbf{(d)}). 
This numerical finding confirms the scenario \ref{pos:1} presented in Sec.~\ref{sec:effectiveTheory}.

\subsubsection{Characterization of the phases: entanglement properties}

\begin{figure}
        \includegraphics[width=\linewidth]{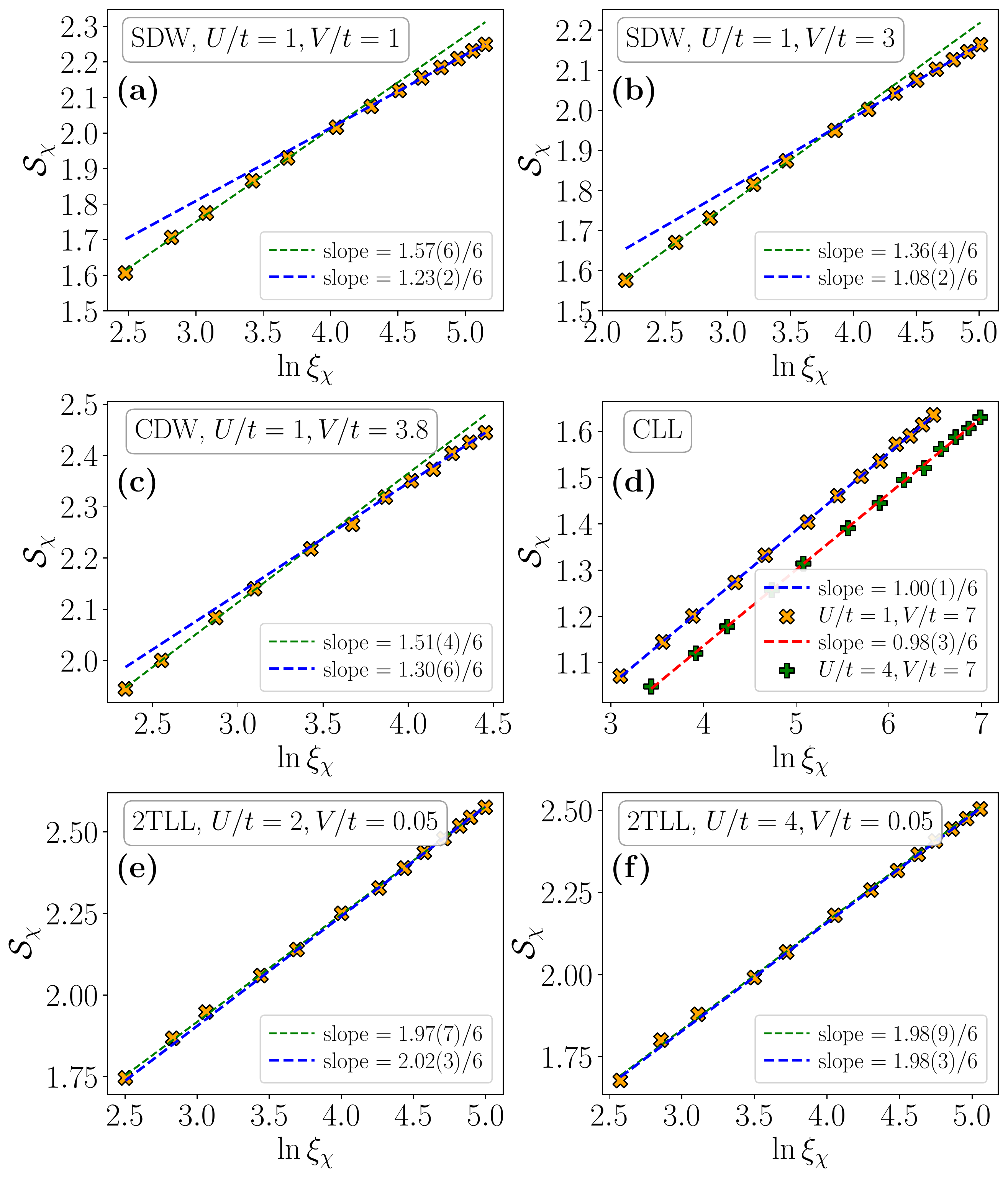}
    \caption{(Color online.) The behavior of the entanglement entropy $\mathcal{S}_{\chi}$ with respect to the correlation length $\xi_{\chi}$ in different phases for the iMPS bond dimension $\chi \in [64, 1280]$ ($\chi \in [96, 1280]$ for CDW phase). \textbf{(a)} and \textbf{(b)} correspond to two points in the SDW phase, \textbf{(c)} and \textbf{(d)} are for the CDW and the CLL phases respectively, while \textbf{(e)} and \textbf{(f)} correspond to the 2TLL phase. In \textbf{(a)}-\textbf{(c)}, \textbf{(e)}, and \textbf{(f)} the thin green dotted lines correspond to the fits in the range $\chi \in [64, 384]$ ($\chi \in [96, 384]$ for the CDW phase in \textbf{(c)}), while the blue dashed lines correspond to the fits for $\chi \in [512, 1280]$.
    }
    \label{fig:central_charge_phase}
\end{figure}

We now characterize the SDW, CDW, CLL, and 2TLL phases that appear in the system for $U/t > 0$ 
by the scaling of the entanglement entropy against the bond dimension, extracting the corresponding central charges.
Specifically, here we show that all these phases -- except the one at weak intra-chain and strong inter-chain coupling -- have $c=1$.

Our theoretical analysis using the Abelian bosonization in Sec.~\ref{sec:weak_coupl_square} tells that the spin sector in the SDW phase is gapped, while the charge sector is not, thereby suggesting that the phase is of $c=1$. The phase is adjacent to a BKT transition that requires too large bond dimensions to be sharply demarcated. Thus, any points in its vicinity cannot be sharply identified by deducing their central charge, unless very large bond dimension is taken that may not be always possible in practical simulations. In Figs.~\ref{fig:central_charge_phase}\textbf{(a)} and \textbf{(b)}, we show the variations of the entanglement entropy $\mathcal{S}_{\chi}$ with the correlation length $\xi_{\chi}$ in two  points in the SDW phase respectively with bond dimensions in the wide range of $[64, 1280]$. By fitting the scaling formula of Eq.~\eqref{eq:cardy_calabrese_infinite} to the data in different ranges of $\chi$, we observe that the slopes of the curves reduce with higher values of $\chi$ and slowly approach to the expected value of $1/6$.

The similar trend in the behavior of the entanglement entropy is observed in the sandwiched CDW phase, where the slope of the curve reduces from $\sim 1.5/6$ to $\sim 1.3/6$ with increasing bond dimension (see Fig.~\ref{fig:central_charge_phase}\textbf{(c)}). This is due to the fact that there are two very nearby transitions with central charge $c > 1$, and we need very large bond dimensions to resolve the CDW phase properly. However, since, in lattice models, quantum phases with only integer values of central charges are stable, the data of Fig.~\ref{fig:central_charge_phase}\textbf{(c)} confirms that the CDW phase is of $c=1$.

On the other hand, deep in the CLL phase with $U/t > 0$ (see Fig.~\ref{fig:central_charge_phase}\textbf{(d)}), we can faithfully obtain the central charge of $c=1$. For the entire range of $\chi \in [64, 1280]$, the slopes of the curves stay stick to $1/6$ in the CLL phase. This  proves that the cluster spin-gap in this CLL phase in non-zero ($K_s < 1$ in Eqs.~\eqref{eq:square_cluster} and \eqref{eq:Cparameters_square}), and the cluster excitations from both the chains are coupled (i.e., a spin-locked CLL). 

Figure~\ref{fig:central_charge_phase}\textbf{(e)} and \textbf{(f)} show the same scaling of entanglement entropy in two points deep in the 2TLL phase. The central charge of $c=2$, as predicted by the Abelian bosonization method, can be faithfully extracted by the scaling for the entire range of bond dimension considered here.

\subsubsection{Correlation functions in the SDW and CDW phases}

\begin{figure}[t]
        \includegraphics[width=\linewidth]{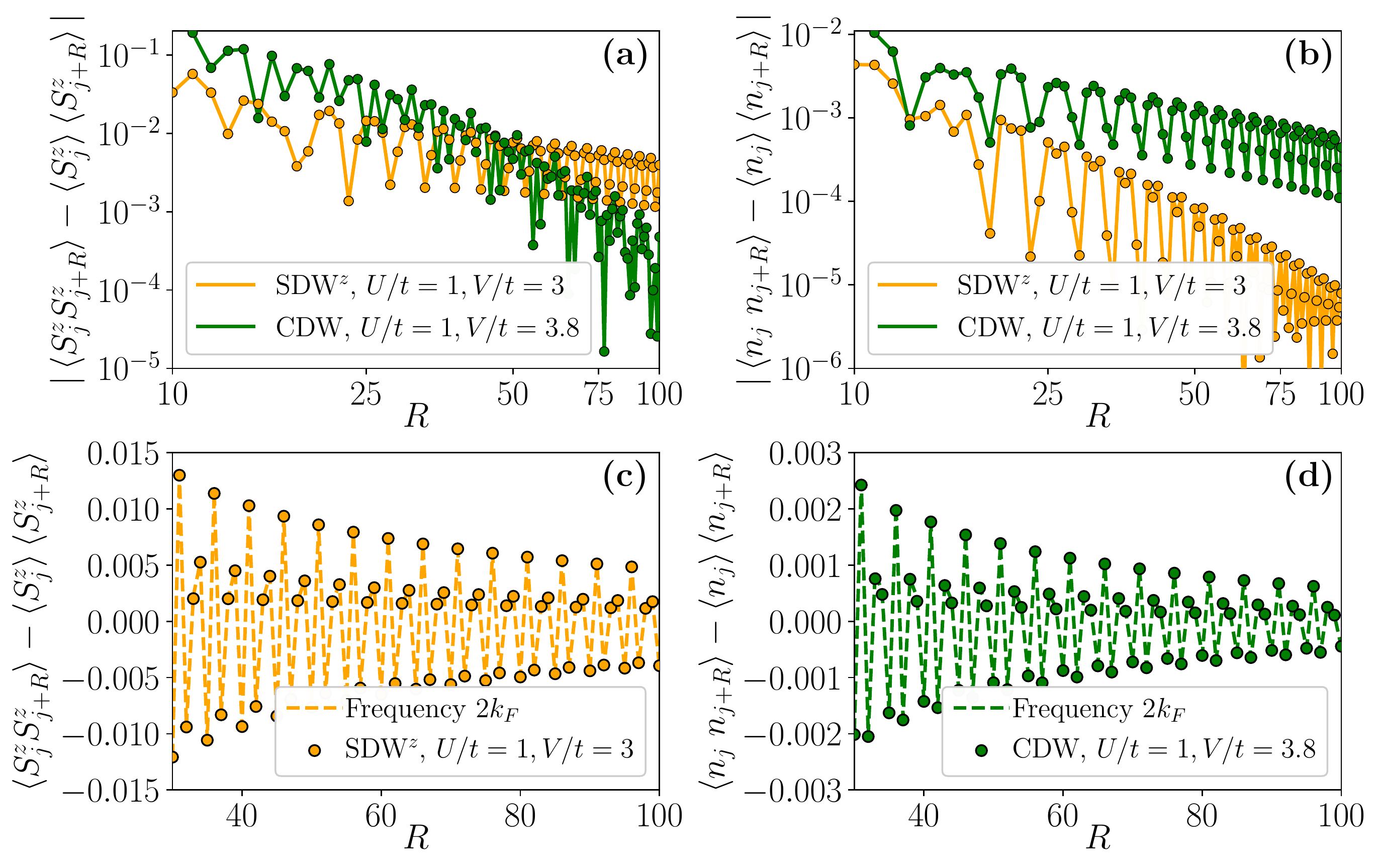}
    \caption{(Color online.) The behaviors of \textbf{(a)} the spin correlation function $\mathcal{C}_{S^z} (R) = \langle S^z_j S^z_{j+R} \rangle - \langle S^z_j \rangle \langle S^z_{j+R} \rangle$ and \textbf{(b)} the charge correlation function $\mathcal{C}_{n} (R) = \langle n_j n_{j+R} \rangle - \langle n_j \rangle \langle n_{j+R} \rangle$ as functions of the distance $R$ 
    in the SDW and CDW phases. Here both axes are in the logarithmic scale. \textbf{(c)}-\textbf{(d)} The fits (according to Eq.~\eqref{eq:power_law_cosine}) of the spin and charge correlations respectively in the SDW and CDW phases.
    }
    \label{fig:correlation_sdw_cdw}
\end{figure}

We now move on to analyze correlation functions (see Eq.~\eqref{eq:correlation_function}) in the SDW and CDW phases. 
Specifically, we first consider the spin correlation function
\begin{equation}
    \mathcal{C}_{S^z}(R) = \langle S^z_j S^z_{j+R} \rangle - \langle S^z_j \rangle \langle S^z_{j+R} \rangle,
\end{equation}
with $S^z_j = (n_{j,+} - n_{j,-})/2$,
and the charge correlation function
\begin{equation}
    \mathcal{C}_{n}(R) = \langle n_j n_{j+R} \rangle - \langle n_j \rangle \langle n_{j+R} \rangle,
\end{equation}
with $n_j = (n_{j,+} + n_{j,-})$,
in the SDW and in the CDW phases. However, as mentioned before, these correlation functions can be faithfully interpreted only when $R \lesssim \xi_{\chi}$, and in these two phases we have $\xi_{\chi} \sim 100$ for the largest bond dimension ($\chi = 1280$) at our disposal.

Figs.~\ref{fig:correlation_sdw_cdw}\textbf{(a)} and \textbf{(b)} depict the correlation functions $\mathcal{C}_{S^z}(R)$ and $\mathcal{C}_{n}(R)$ in the SDW and CDW phases for $R \leq 100$. We remark that for this small range of $R$ differentiating the power-law decay with the exponential decay is possible only under the assumption of a correlation length shorter that $\xi_{\chi}$. Still, the trends of correlations in Fig.~\ref{fig:correlation_sdw_cdw} are quite distinctive from each other. Specifically, we observe slow power-law decay of the spin correlation and fast (exponential) decay of the charge correlation in the SDW phase, numerically confirming the spin-density wave nature of the phase. Exactly opposite is seen in the CDW phase, that has oscillating slow (power-law) decay of the charge correlation. This provides convincing proof that this sandwiched phase is indeed a CDW.

To correctly get the proper frequencies of the oscillations in the algebraically decaying correlations, we numerically fit them using the following formula:
\begin{equation}
    \mathcal{C}(R) \sim \cos\left(k R \right) R^{-\beta}.
    \label{eq:power_law_cosine}
\end{equation}
In Figs.~\ref{fig:correlation_sdw_cdw}\textbf{(c)} and \textbf{(d)} we show such fits for the spin and the charge correlations respectively in the SDW and CDW phases. In both the cases, the numerical fits show that the frequency of oscillations are $k = 2 k_F$ as expected from Eqs.~\eqref{eq:SDW_spin},~\eqref{eq:CDW_both}, and~\eqref{eq:SDW_charge}.

\subsubsection{Correlation functions in the spin-locked CLL phase}

\begin{figure}
    \centering
    \includegraphics[width=\linewidth]{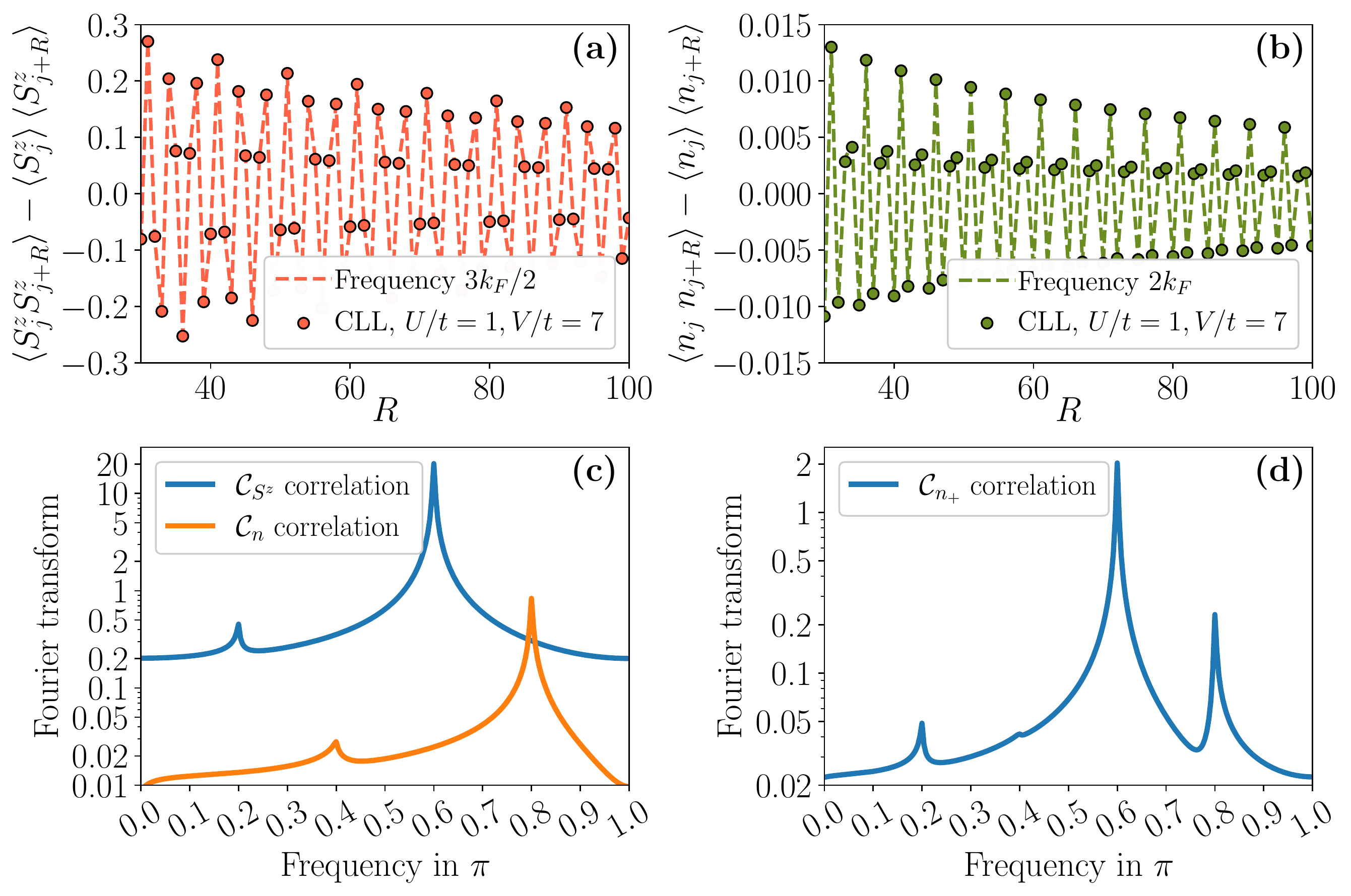}
    \caption{(Color online.) \textbf{(a)} The spin correlation  $\mathcal{C}_{S^z} (R)$ and \textbf{(b)} the charge correlation $\mathcal{C}_{n} (R)$ as functions of the distance $R$ 
    in the CLL phase. The dashed lines show the numerical fits according to Eq.~\eqref{eq:power_law_cosine}.
    The subplot \textbf{(c)} depicts the Fourier transform of the correlations $\mathcal{C}_{S^z} (R)$ and $\mathcal{C}_{n} (R)$, while \textbf{(d)} shows the same for the correlation function of the single-chain density $n_{+}$.
    }
    \label{fig:correlation_cll}
\end{figure}

By analyzing the spin and charge correlation functions in the spin-locked CLL phase, we observe that, unlike the SDW and CDW phases, both the correlations follow  power-law behavior (see Figs.~\ref{fig:correlation_cll}\textbf{(a)} and \textbf{(b)}). Both the numerical fits using Eq.~\eqref{eq:power_law_cosine} and their Fourier transformation Fig.~\ref{fig:correlation_cll}\textbf{(c)} suggests that the frequencies of the oscillations are $k=3 k_F /2 = 3 \pi/5$ for the spin correlation function and $k = 2 k_F = 4 \pi / 5$ for the charge correlation. $k=3 k_F /2$ is the cluster Fourier momentum as it appears in Eq.~\eqref{density-cbosonization} and translates the specific cluster instability of the Luttinger liquid. Surprisingly, this feature is absent in the charge correlation function, and is instead replaced by a peak at $2 k_F$. Since we lack a controlled operator mapping between lattice and underlying fields, it is not clear what is the origin of such property. Still, we point out that single chain correlations do show a strong peak at the clustering point. 
Indeed, the correlation function for the single-chain density $n_{\ell}, \ell=\pm$, is
\begin{equation}
    \mathcal{C}_{n_{\ell}} (R) = \langle n_{j, \ell} n_{j+R, {\ell}} \rangle - \langle n_{j, {\ell}} \rangle \langle n_{j+R, {\ell}} \rangle.
\end{equation}
Fig.~\ref{fig:correlation_cll}\textbf{(d)} shows the Fourier transform of the correlation function $\mathcal{C}_{n_+}$. This Fourier transform shows both -- a peak at the frequencies $k = 3 \pi /5$ corresponding to the cluster instability and $k = 4 \pi /5$ corresponding to the possible rivalling density wave instability. This observation suggests that one possible explanation for the lack of cluster peaks in the CDW correlator is that inter-leg correlations are out of phase with respect to intra-leg ones.

\subsection{The complete picture from the numerical analysis}

The above numerical investigations using both ED and TN methods provide us the full picture about the system phase diagram and different phases. Here we summarize the details of the phase diagram with increasing $V/t$ in the following (compare the schematic phase-diagram in Fig.~\ref{fig:Ryd}\textbf{(b)}):
\begin{enumerate}
    \item At small values of $V/t \sim 0$, the ground state of the system is in standard 2TLL phase (with $c=2$).
    \item With increasing $V/t$, a BKT transition appears and 
    the system goes into the $c=1$ SDW phase where quasi-long-range order in the spin sector exists.
    In the phase, the spin sector becomes gapped while the charge sector remains gapless. The precise determination of this BKT transition, predicted by Abelian bosonization in the weak coupling regime, is beyond the scope of the numerical tools used here.
    
    \item There exists a $c=2$ Gaussian transition that separates the SDW phase from the CDW phase. Across this transition the spin-gap closes, so that both the charge and the spin sectors are gapless resulting in $c=2$.
    \item In the CDW phase, the spin sector again becomes massive with disordered paramagnetic nature, while the charge sector remains gapless with a charge density wave order giving us $c=1$.
    \item After the CDW phase, a $c=3/2$ phase transition appears, where one part $c=1$ is again from the gapless charge sector, and the another part $c=1/2$ comes from the Ising type transition same as in the single-chain SUSY point.
    \item At high values of $V/t$ and $U/t >0$, a $c=1$ CLL phase exists. However, unlike the 2CLL phase (with $c=2$) at $U/t=0$, the clusters in both the chains are coupled, and the cluster spin-gap is non-vanishing that results into $c=1$. 
\end{enumerate}

\section{Conclusion and discussions}
\label{sec:conclu}

In this work, we provided and characterized the phase diagram of a system of spinless hard-core bosons in a square ladder lattice at filling $\nu = 2/5$, interacting via soft-shoulder potentials, and in the absence of inter-chain tunneling. These models  are directly inspired by experimental setups using Rydberg atoms loaded in an optical lattice~\cite{Guardado-Sanchez2021}, where most of the regimes we propose are in principle  accessible. 

To obtain the phase diagram, we used analytical arguments close to the regime of weak coupling between the legs ($U\ll t, V$). Our analytical predictions were corroborated and extended to larger coupling using numerical simulations. Specifically, for $V\ll t$ and $U=0$, each chain is described by a gapless Tomonaga-Luttinger liquid (TLL) with central charge $c=1$. As soon as $U>0$, the degree of freedom relative to the leg of the ladder, called spin, becomes  gapped. The resulting gapless phase displays a leading spin-density wave (SDW) instability of momentum $2k_F$ well seen in the simulations. When $V\lesssim U/5$, and particularly when $V=0$, the system 
transits across a Berezinskii-Kosterlitz-Thouless (BKT) phase transition predicted by the Abelian bosonization and hinted by the exact diagonalization. Past the transition, the description of the system is equivalent to the standard Hubbard model described by two Luttinger liquids with $c=2$ in total. 

For $V\gg t$ and $U=0$, each chain displays an exotic cluster Luttinger liquid (CLL) with $c=1$ described by cluster bosonization and seen by iDMRG.
As soon as $U>0$, the cluster equivalent of the spin degree of freedom is gapped, leading to a `spin-locked' cluster with $c=1$. The corresponding cluster density wave has a momentum $(3/4)\times 2k_F$, the clearest consequence of the clusterization of the microscopic degree of freedom into cells of 3 clusters of 4 particles in total for this filling and the range of $V$. Such emergent clustering is also reflected in both single chain and spin correlation functions.

In the vicinity of the single-chain transition point, we develop an effective field theory describing the ladder system that is obtained via coupling two supersymmetric conformal field theories -- a scenario that is, to the best of our knowledge, not accessible in any other cold atom setup. 
The SUSY phase transition point at $U=0$ and $V/t\simeq 5.7$ extends into a gapless phase with $c=1$ and leading charge-density wave instability (CDW) of momentum $2k_F$ when $U>0$ and $V/t\sim 5.7$. The CDW is separated from the SDW by a Gaussian phase transition with $c=2$ and from the spin-locked CLL by a critical line with $c=3/2$. Both the phase and the transitions are sharply seen by iDMRG. The critical line $c=3/2$ may also be supersymmetric conformal phase transitions, but it is not possible to conclude this based on the present analysis. In fact, such proof may require a true quantum simulator, as the more direct way of probing this is studying the long-time dynamics of large chains. Indeed, it is possible to obtain the low-energy band structure of time-independent, translation-symmetric Hamiltonian by computing the time and space Fourier transform of correlation functions of various observables after a quench~\cite{Villa2019,Villa2020}. The system therefore provides a unique platform to study interactions between two SUSY conformal field theories.

Beyond observability in experiments, that is certainly the main drive behind our research line, there are several research directions that can be pursued starting from the understanding of the present model.
The onset of CLLs require a range of soft-shoulder interaction of at least 2. They are generally not modified by longer range, so that we do not expect a qualitative change of the phase diagram for longer ranges. Other commensurate fillings (above $1/3$ but below $1/2$ for range 2) may display CLL physics leading to similar phase diagrams. Different fillings for both chains are instead likely to lead to richer arrangements between the CLL on the two legs and decouple the SUSY points on one leg and the other, enriching the phase diagram once more. Other CLL arrangements may also appear when changing the sign of $U$, or extending the interaction $U$ over more rungs.
Besides exploring these various regimes theoretically and experimentally, it would also be interesting to stack more chains above the ladder and approach the 2D regime. It is indeed not known if the clusterization vanishes or if it becomes the local order of a fully gapped phase and how. Finally, it would be interesting to investigate regimes where the ladder displays a genuine SU(2) ``spin"-symmetry, in analogy to electronic systems. Such interactions are not available when dressing atoms via $s$-states, but could be at least approximately realized utilizing a combination of dressing to $p$-states~\cite{Alex2015} and lattice spacing tuning.

\begin{acknowledgments}
We acknowledge useful discussions with A. Angelone, M. Fabrizio, G. Giudici, G. Japaridze and P. Lecheminant. The work of M. D., P. F. and M. T. is partly supported by the ERC under grant number 758329 (AGEnTh), by the MIUR Programme FARE (MEPH), and by the European Union's Horizon 2020 research and innovation programme under grant agreement No 817482 (Pasquans). M. T. thanks the Simons Foundation for supporting his Ph.D. studies through Award 284558FY19 to the ICTP. P.F. thanks Pr. Dr. Loss and Klinovaja and the University of Basel for the hospitality during the writing of the manuscript. The work by A.N. was partly supported by the Shota Rustaveli National Science Foundation of
Georgia, SRNSF, Grant No. FR-19-11872. The iDMRG simulations have been performed using the TeNPy library \cite{tenpy}. The code used to obtain the data presented in the paper can be found in~\cite{codes}.
\end{acknowledgments}

\appendix*

\section{Long-range intra-leg interaction potential $V(\mathbf{r})$ at weak coupling}
\label{app1}

Using the Abelian bosonization approach discussed in Sec.~\ref{sec:weak_coupl_square}, the weak coupling effective field theory is derived for a generic case of the intra-leg soft-shoulder potential range $r_C$. Away from half-filling, the bosonized Hamiltonian has exactly the same form as Eq.~\eqref{eq:boso}. The charge sector is gapless. The bare parameters describing the spin sector at weak coupling are
\begingroup
\allowdisplaybreaks
\begin{subequations}\label{eq:parameters_app}
\begin{align}
    u_s K_s &= u_c K_c = v_F,\\
    K_s &= \frac{1}{\sqrt{1+\frac{g_{\parallel}}{2\pi v_F}}},
    \\
    g_\parallel &= -2\left(U-V\left(2r_c+1-\frac{\sin([2r_c+1]k_F)}{\sin(k_F)}\right)\right),
    \\ 
    K_c &= \frac{1}{\sqrt{1-\frac{g}{2\pi v_F}}},
    \\
    g &= -2\left(U+V\left(2r_C+1-\frac{\sin([2r_C+1]k_F)}{\sin(k_F)}\right)\right).
\end{align}
\end{subequations}
\endgroup
while $g_\perp=-2U$ remains unaffected. For $k_F=2\pi/5$, the changes in parameters Eqs.~\eqref{eq:parameters_app} narrows the location of the BKT phase transitions for $V<U/5$ but leaves the first loop prediction ($V=0$) unchanged.

\bibliography{manuscript.bbl}

\end{document}